\documentclass[twocolumn]{aastex63}
\pdfoutput=1 

\usepackage[colorinlistoftodos]{todonotes}
\journalinfo{ }

\let\Oldtodo\todo
\renewcommand{\todo}[1]{\Oldtodo[inline]{#1}}

\newcommand{\addtxt}[1]{#1} 
\newcommand{\blue}[1]{#1} 
\newcommand{\rev}[1]{#1} 

\usepackage{nicefrac}

\usepackage{xspace}

\usepackage{amssymb}
\usepackage{amsmath}
\usepackage{lineno}
\usepackage{appendix}

\newcommand{\swift}{\textit{Swift}\xspace}
\newcommand{\rosat}{\textit{ROSAT}\xspace}
\newcommand{\xmm}{\textit{XMM}\xspace}
\newcommand{\chandra}{\textit{Chandra}\xspace}

\begin{document}

\title{An Optical Search for \blue{New} Outbursting Low Mass X-Ray Binaries}

\author[0000-0001-5538-0395]{Yuankun Wang}

\author[0000-0001-8018-5348]{Eric C. Bellm}

\author[0009-0000-6087-1375]{Allison Crossland}
\affiliation{DIRAC Institute, Department of Astronomy, University of Washington, 3910 15th Avenue NE, Seattle, WA 98195, USA}

\author[0000-0002-2577-8885]{William I. Clarkson}
\affiliation{Natural Sciences Department, University of Michigan-Dearborn, 4901 Evergreen Road, Dearborn, MI 48128, USA}

\author[0000-0002-7503-5078]{Alessandro Mazzi}
\affiliation{Dipartimento di Fisica e Astronomia Galileo Galilei, Università di Padova , Vicolo dell’Osservatorio 3, I-35122 Padova, Italy}


\author[0000-0002-0387-370X]{Reed Riddle}
\affiliation{IPAC, California Institute of Technology, 1200 E. California Blvd, Pasadena, CA 91125, USA}

\author[0000-0003-2451-5482]{Russ R. Laher}
\affiliation{IPAC, California Institute of Technology, 1200 E. California
             Blvd, Pasadena, CA 91125, USA}

\author[0000-0001-7648-4142]{Ben Rusholme}
\affiliation{IPAC, California Institute of Technology, 1200 E. California
             Blvd, Pasadena, CA 91125, USA}

\begin{abstract}
Transient Low-Mass X-ray binaries (LMXBs) are discovered largely by X-ray and gamma-ray all-sky monitors. The X-ray outburst is also accompanied by an optical brightening, which empirically can precede detection of X-rays.  Newly sensitive optical synoptic surveys may offer a complementary pathway for discovery, and potential for insight into the initial onset and propagation of the thermal instability that leads to the ionization of the accretion disk. We use the Zwicky Transient Facility (ZTF) alert stream to perform a comprehensive search at optical wavelengths for \blue{previously undiscovered} outbursting LMXBs. Our pipeline first crossmatches the positions of the alerts to cataloged X-ray sources, and then analyzes the 30-day lightcurve of matched alerts by thresholding on differences with an 8-day exponentially weighted moving average.  In addition to an nineteen month-long live search, we ran our pipeline over four years of ZTF archival data, recovering \rev{4} known LMXBs. We also independently detected an outburst of MAXI J1957+032 in the live search and found the first outburst of Swift J1943.4+0228, an unclassified X-ray transient, in 10 years. Using Monte Carlo simulations of the Galactic LMXB population, we estimate that ~29\% of outbursting LMXBs are detectable by ZTF and that ~4.4\% of LMXBs would be present in the crossmatched X-ray catalogs, giving an estimated Galactic population of \addtxt{$3390^{+3980}_{-1930}$}. We estimate that our current pipeline can detect 1.3\% of \rev{all outbursting LMXBs, including those previously unknown,}  but that Rubin Observatory's Legacy Survey of Space and Time (LSST) will be able to detect 43\% of outbursting LMXBs.
\end{abstract}


\keywords{LMXBs, X-ray binaries, sky surveys}



\section{Introduction}
X-ray binaries (XRBs) are interacting binary systems containing a neutron star or black hole and a stellar companion. 
In Low Mass X-ray Binaries (LMXBs), \addtxt{matter from a low mass stellar companion is transferred into an accretion disk via Roche Lobe overflow. The} accretion of matter onto the compact object does not happen uniformly, which leads the LMXB to undergo outbursts \citep{Shakura:1973, King:1998:SoftX-rayLCs}. 
The most widely discussed mechanism for these outbursts is the Disk Instability Model \citep[DIM; for a detailed review, see][]{Ham:2020:DIMReview}. 
In this model, the luminosity of these systems can increase by many orders of magnitude when an instability \citep[perhaps the Magnetorotational Instability;][]{Balbus:1998:InstabilityInACs} causes the accretion disk to become ionized, leading the mass transfer rate onto the compact binary to dramatically increase and causing \addtxt{a transient} outburst event. \addtxt{High Mass X-ray Binaries (HMXBs) are mostly persistent, wind-fed systems. Many, but not all, of these objects lack accretion disks and prominent optical outbursts \citep{Reig:2015:OptHMXBs}.}

\addtxt{Accretion} outbursts of \addtxt{LMXBs} typically exhibit a characteristic fast rise, exponential decay temporal profile \citep{Chen:1997}. 
The initial rise of several days to weeks corresponds to complete ionization of the accretion disk and increase in mass transfer onto the compact object, followed by a long decay as energy is radiated, cooling the ionized disk. 
The decay portion of the outburst can last for hundreds of days or even years, as irradiation from the compact object can sustain ionization of the inner disk \citep{King:1998:SoftX-rayLCs, Dubus:2001:DIMinXRBs}. 
The end of the outburst can transition to a linear decay \citep{King:1998:SoftX-rayLCs, Tetarenko:2018:XrayIrr} as the disk changes back to its neutral state. 
While the narrative provided by the disk instability model (DIM) matches the overall observed behavior of outbursts, many key details of the process remain uncertain, including the mechanism that triggers the outburst onset. The location where this instability originates and the process by which it propagates throughout the disk are also not well understood.
Also uncertain are why some outbursts fail to progress to the disk-dominated high-soft state, resulting in a smaller and shorter ``failed transition'' outbursts \citep{Alabarta:2021:FailedOutburst}. 

Many modifications to the model have been proposed to account for observations, including irradiation of the disk \citep{King:1998:Irr, Niwano:2023:aqlx1}, tidal forces \citep[e.g.,][]{Priedhorsky:1988:TidalXRBs}, and disk winds  \citep[e.g.,][and references therein]{tmp_Neilsen:23:XRBWindReview}.
Discoveries of both new XRBs and new XRB outbursts may help answer these questions and clarify the underlying physical processes that necessitate these modifications.

The discovery of X-ray Binaries has largely been driven by X-ray all-sky monitors, such as the All-Sky Monitor (ASM) on the Rossi X-ray Timing Explorer (RXTE). However, while the sensitivity of these instruments, around $10^{-8}\mathrm{\,ergs\,cm^{-2}\,s^{-1}}$ (about $10^{35}-10^{36}$\,ergs\,s$^{-1}$ at fiducial galactic distance \citep{Krimm:2013:Swift_BAT, Matsuoka:2009:MAXI}), can detect the intense radiation from outbursts, they miss X-ray Binaries in quiescence ($10^{29}-10^{33.5} \mathrm{\,ergs\,s^{-1}}$), and can miss weak/low luminosity outbursts,  \addtxt{such as outbursts from very faint x-ray transients \blue{\citep[VFXTs; e.g.,][]{Padilla:2014:FaintestBH,Heinke:2015:VFXTs}}}. Focusing X-ray telescopes such as Chandra, XMM-Newton, and Swift have the sensitivity to detect much fainter sources and their higher resolution also provides more precise spatial localization. However, their smaller field of view means that their combined sky coverage is much smaller than the all-sky monitors. Additionally, classification of detected X-ray radiation as an outbursting X-ray Binary can be difficult with both all sky monitors and the more sensitive X-ray source catalogs, particularly for fainter, shorter duration outbursts that are only observed serendipitously, meaning that a number of LMXBs may be catalogued but not classified.

Optical detection and monitoring offers an alternate pathway to discovery and characterization of X-ray binaries. Empirically, the initial stages of the outburst can first be observed at optical wavelengths several days before the X-ray outburst \citep{Rus19}. For example, the X-ray Binary AT2019wey \citep{Yao21} was discovered first in the optical by ATLAS \citep{TNS:2019:ATLASdet2019wey} and was monitored by the Zwicky Transient Facility months before a detection of an X-ray outburst \citep{Atel:2020:SRGdetAT2019wey}. Since the optical outburst can precede the X-ray brightening, this can enable earlier X-ray followup during the initial rise and the study of the mechanisms that trigger the outburst, which are not understood due to a relative rarity of observations during the earliest phases of outburst onset. However, an optical search requires a fast cadence to identify candidates in the short brightening phase, while also needing the depth to catch faint, extincted sources in the Galactic Plane. \addtxt{In addition, optical fields are much more crowded at typical outburst magnitudes compared to the X-ray sky and X-ray outbursts. As a result, additional contextual data, such as the lightcurve history, and prompt followup are needed to identify outbursts as LMXBs early in the outburst cycle.}

\addtxt{
The X-ray Binary New Early Warning System (XB-NEWS) collaboration \citep{Rus19} uses observations from the Las Cumbres Observatory to monitor known X-ray binaries and has made the first detection of several LMXB outbursts \citep{XBnews:2020:detMAXIJ1348, Baglio:2022:CenX4_failed_ob, Saikia:2023:OptMonBXBH}. 
}
\blue{This dedicated program ensures consistent monitoring of known XRBs over long periods, but requires dedicated observations and data processing.  Additionally, its narrow-field targeted observations are not able to discover outbursts from previously unknown XRBs.}

\blue{Optical synoptic surveys provide an alternative means of discovering outburts from both known and new LMXBs.}
The Zwicky Transient Facility \citep[ZTF;][]{Bellm:2019:ZTFsystem, Graham:2019:ZTFscience, Dekany:2020:ObservingSystem} images the northern sky down to $\sim$21 mag at a cadence of 1--3 days \citep{Bellm:2019:Scheduler}. The high cadence and depth of ZTF combined with its real-time alert system \citep{Pat19}, enables multiple detections of LMXBs during the fast rise phase of outburst and quick investigation of the candidate sources, and motivates this study to find optical outbursts of LMXBs. The easily accessible lightcurve history in ZTF also enables us to eliminate a large number of contaminant sources, such as cataclysmic variables (CVs), by examining the historical outburst behavior and duty cycle of candidate sources.

\blue{For identification of outbursts from known LMXBs in ZTF, 
we make use of a watchlist to notify us of any activity from positions of sources in the Ritter and Kolb \citet{Ritter:2003:XRBcat}, BlackCAT \citep{Corral-Santana:2016:BlackCAT}, and WATCHDOG \citep{Tet16} catalogs. This watchlist, implemented using the ANTARES broker \citep{Matheson:21:ANTARES}, led to the discovery of outbursts from XTE J1859$+$226 \citep{Bellm:2021:ZTF21aagyzqr, tmp_Bellm:23:XTEJ1859} and Swift J1357.2$-$0933 \citep{Bellm:21:SwiftJ1357ATel}.}

In this work, we \blue{focus on searches for previously-unknown LMXBs using the ZTF alert stream.  We describe a live nightly outburst search pipeline which crossmatches ZTF alerts with catalogs of X-ray sources to narrow down the number of candidates considered. We also perform this search over 4 years of archival alert data.} In section \ref{sec:methods}, we describe our methodology for identifying candidates. In section \ref{sec:results}, we present the results of our search, which includes recoveries of known LMXB outbursts over this timespan. In section \ref{sec:results}, we estimate the percentage of galactic LMXBs are visible to our pipeline. In section \ref{sec:discussion}, we conclude with a discussion of our pipeline, and the implications of our results for discovery of LMXBs in future surveys, such as the Vera C. Rubin Observatory's Legacy Survey of Space and Time \citep[LSST;][]{Ivezic:2019:LSST}.

\section{Methods\label{sec:methods}}

\subsection{The ZTF Alert Stream}
The alert stream is a data product provided by the Zwicky Transient Facility. An automated pipeline detects transient and variable sources by differencing nightly images and coadded reference images and identifying any sources that appear in the difference images\footnote{\url{https://irsa.ipac.caltech.edu/data/ZTF/docs/ztf_explanatory_supplement.pdf}} above a SN $>$ 5 \citep{Masci:2019:ZTFpipeline}. Candidates are packaged with metadata, detection history, and image cutouts into alerts. The ZTF alert distribution system \citep[ZADS;][]{Pat19} uses Kafka to stream these alerts out in near real time, enabling rapid followup after identification. The median number of ZTF alerts per observing night for all programs is 363,000, but is highly variable, and can exceed 1M in a single night when observing in the Galactic plane.

Galactic variables frequently appear in both reference and science images, so to find the total apparent magnitude we add the reference and difference fluxes. The apparent photometric magnitudes for each source are determined by adding (or subtracting, in the case of a negative source in the difference image) the PSF-fitted photometry to the reference image photometry. Sources that are too close in brightness to the reference image may not pass the 5-sigma threshold for triggering an alert and show up in the alert history as a nondetection. We determine the limit of nondetections by averaging the upper and lower bounds for the nondetection, which are found by adding and subtracting the difference image limiting magnitude from the reference magnitude. Adding nondetections to the lightcurve history gives greater temporal coverage to the outburst identification algorithm and provides a level of comparison between the outburst and quiescent states.

\begin{figure*}[ht]
\begin{center}
\includegraphics[width=0.9\textwidth]{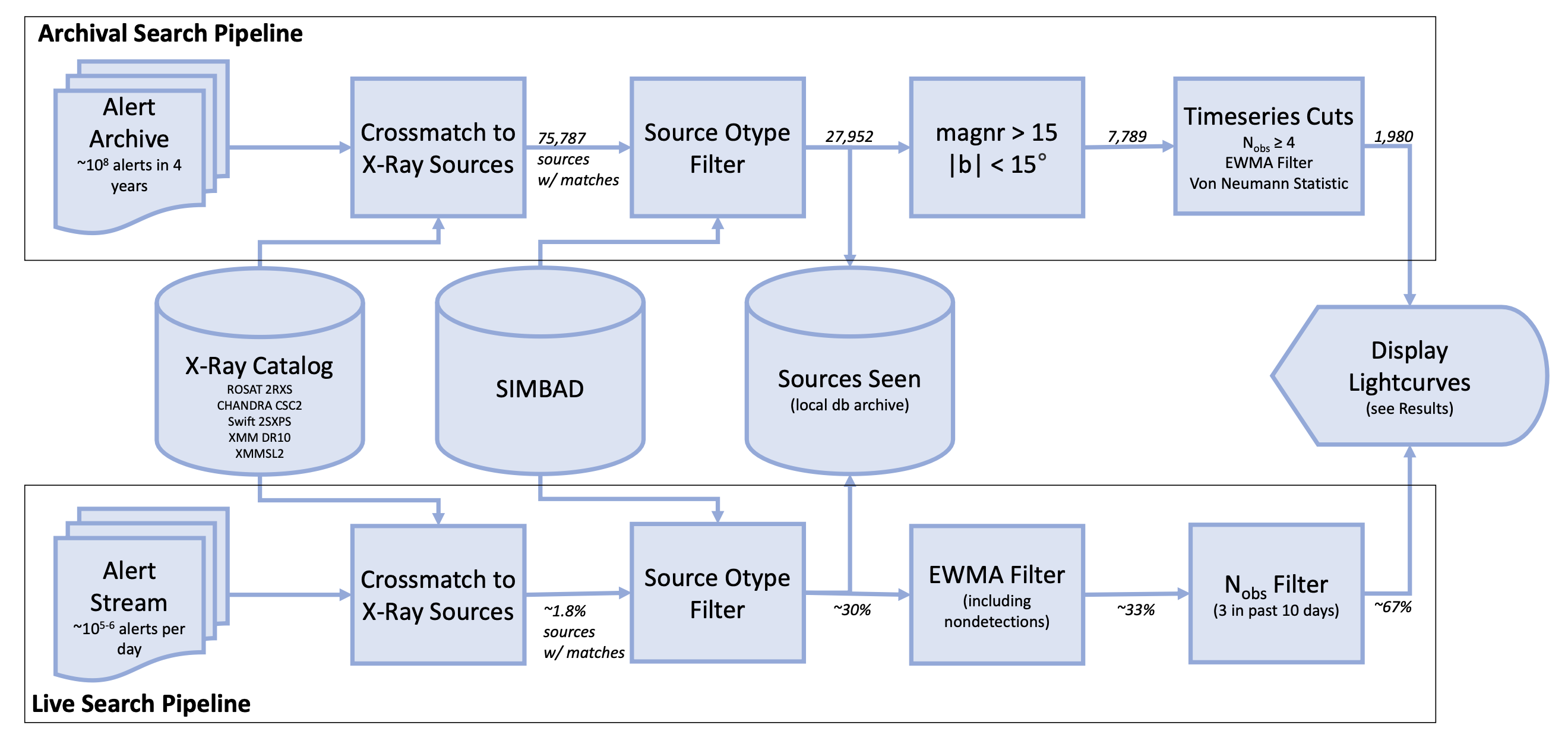}
\caption{Flowchart showing how packets from the alert stream or alert archive are processed by our pipeline. We examine lightcurves of candidates passing all cuts by eye and further investigate their historical observations by ZTF and other surveys.
\label{fig:pipeline}}
\end{center}
\end{figure*}

\subsection{Live Search Pipeline}

\subsubsection{Consuming Alerts} \label{consume-alerts}
\addtxt{The live search over all alerts in the alert stream covered in this paper begins on February 24th, 2021 and ends September 21st, 2022.} ZTF provides access to alerts through a number of community brokers\footnote{e.g., ANTARES \citep{Matheson:21:ANTARES}; Lasair \citep{Smith:19:Lasair}; ALeRCE \citep{Forster:21:ALeRCE}; AMPEL \citep{Nordin:19:AMPEL}; Fink \citep{Moller:21:FINK}; Pitt/Google (\url{http://pitt-broker.readthedocs.io})} which consume the public stream and enable users to filter, query, and otherwise interact with the alerts. However, we consume the alert stream directly from the ZADS system for greater flexibility and control over the filters and cuts in our search for LMXBs\footnote{\url{https://github.com/dirac-institute/alert_stream_crossmatch}}. The filters and cuts on the alert stream are diagrammed in figure \ref{fig:pipeline}.

We eliminate sources with a deep real bogus score \citep{Duev:2019:drb} under 0.8, as well as any packets with nearby known solar system objects. The sources passing this initial cut are then crossmatched to positions of known X-ray sources.

\subsubsection{X-ray and SIMBAD Crossmatch}
In order to narrow down the number of candidates considered, we first crossmatch all alerts with positions of previously catalogued X-ray sources using the given spatial uncertainties in the catalogs as the crossmatch radius. These sources are compiled from 5 X-ray surveys listed in table \ref{tab:cat}: ROSAT 2RXS \citep{Boller:16:2RXS}, Chandra CSC2 \citep{Evans:10:CSC}, Swift 2SXPS \citep{Evans:20:2SXPS}, XMM DR10 \citep{Webb:20:4XMMDR10}, and XMMSL2 \citep{Warwick:12:XMMSlewSurvey}. 
We crossmatched the X-ray source positions to SIMBAD to remove known extra-galactic sources, such as AGN, as well as other previously-identified X-ray emitting sources that are not X-ray Binaries. We then created a unified X-ray dataset combining all 5 catalogs. The ROSAT 2RXS and XMMSL catalogs provide comprehensive sky coverage at the expense of poor localization, and roughly 15\% of sources in these catalogs overlap with better localized sources in the Chandra and Swift catalogs.  The total number of X-ray sources we crossmatch against is 848651.

\begin{table*}
    \begin{tabular}{|l|l|l|r|r|r|}
        \hline
        Observatory & Catalog & Energy & Median $1\sigma$ & Sky & Number \\ %
        & & Range & Localization & Coverage & of Sources \\ 
        \hline
        \rosat & 2RXS \citep{Boller:16:2RXS} & 0.1--2.4\,keV & 15.7\arcsec
        & 100\% & 135,118 \\
        \xmm & XMMSL2 \citep{Warwick:12:XMMSlewSurvey} & 0.2--12\,keV & 4.2\arcsec & 84\% & 29,393 \\
        \swift-XRT & 2SXPS \citep{Evans:20:2SXPS} & 0.3--10\,keV & 2.7\arcsec & 9.1\% & 146,768\\ 
        \xmm & 4XMM-DR10 \citep{Webb:20:4XMMDR10} & 0.2--12\,keV & 1.5\arcsec & 3.0\% & 575,158 \\ 
        \chandra & CSC 2.0 \citep{Evans:10:CSC} & 0.1--10\,keV & 0.6\arcsec & 1.3\% & 317,167 \\
        \hline
    \end{tabular}
\caption{X-ray catalogs used for crossmatching in this work.}
\label{tab:cat}
\end{table*}

ZTF alerts within the positional uncertainty of an X-ray source are flagged as matches and saved for further analysis. Since the positional uncertainty of candidates from ZTF is usually smaller than their matched X-ray counterparts, we again crossmatch to SIMBAD to eliminate any unwanted contaminants. Sources passing the SIMBAD cut are saved into a SQLite database along with photometric and nondetection data.

\subsubsection{Time Series Metrics} \label{sec:ts_metrics}

We use photometric magnitudes to calculate our timeseries features. Objects with reference magnitudes brighter than 15 mag are eliminated, since these sources are near ZTF's saturation limit and are unlikely to be undiscovered LMXBs. Additionally, we eliminate the 0.5\% most uncertain observations (photometric errors greater than 0.5\,mag), which are usually the result of problems in the data quality. This eliminates a number of spurious observations.

We then calculate an 8-day ($\tau=8$) exponentially weighted moving average (EMA) on the apparent magnitudes and nondetection limits: 
\begin{equation} \label{eq:EMA}
\begin{aligned}
\begin{gathered}
    EMA(t) = f * EMA(t_p) + (1-f) * mag(t) \\
    f = \exp{-(t-t_p)/\tau}    
\end{gathered}
\end{aligned}
\end{equation}
Where $t$ is the time of detection and $tp$ is the time of the previous detection. The EMA is a trailing metric: during an outburst, the photometric magnitude will be brighter than the EMA at the same time, leading to a positive difference between the 8-day EMA and the photometric magnitude. For an outburst, we expect this difference to be positive for at least 8 days. We use a 4-day rolling average of this EMA difference as our metric for outburst, selecting all objects where this metric exceeds 0.25 for at least one observation. We then sort all candidates from the previous 10 days by the maximum value of this metric and visually inspect the lightcurves. 

The EMA cut typically provides on order 100 lightcurves per day. We examine the complete ZTF photometric history of lightcurves passing the EMA cut. This allows us to eliminate a large number of cataclysmic variables, which are by far the most common contaminant. The recurrence time and outburst duration of CVs \citep{Coppejans:2016:CVstats} are both generally much shorter than those for LMXBs \citep{Tet16}. Additionally, CV outbursts, which are much less luminous than LMXB outbursts, tend to be closer and therefore much less extincted (i.e red) than LMXB outbursts.

Candidates exhibiting a lack of outburst history or long-duration outbursts with long recurrence times are examined further using the ZTF forced photometry service. We also check for \addtxt{past observations in X-ray by \textit{Swift} XRT and BAT, in SIMBAD for publications and references, and for observations and spectra by other surveys such as SDSS}.

We follow up sources by triggering spectroscopic observations in the optical using the 3.5m telescope at Apache Point Observatory (APO), as well as in the X-ray using a \textit{Swift} ToO. The spectroscopic followup enables us to identify the presence of double-peaked Balmer emission lines. While these lines are also present in the spectra of cataclysmic variables (CVs), black hole binaries have broader emission lines than CVs at a given orbital period \citep{Casares:2018:PhotXRBs}. The full width half maximum (FWHM) of these emission lines can both serve as a diagnostic and be used to estimate the mass ratio of the compact object to the companion star \citep{Casares:2016:MassRatioHa}.

\subsection{Archival search}
In addition, we also ran a search for outbursting LMXBs over 4 years of archived ZTF alert data. Similar to the live search, archived alerts were first crossmatched against our combined X-ray catalog, with matches then crossmatched to SIMBAD to eliminate extragalactic and other contaminant sources. We then implemented the same cuts on reference magnitude and observation uncertainty. We also required each candidate to have at least 4 observations in a single filter over a stretch of 28 days.

Candidates that passed these basic cuts were then subjected to cuts based on timeseries features. We again calculate the EMA metric described in \ref{sec:ts_metrics}, as well as the Von Neumann statistic:

\begin{equation}\label{eq:VN}
    \eta = \frac{1}{(N - 1)\sigma^2}\sum_{i=1}^{N-1}(m_{i+1}-m_{i})^2
\end{equation}

The Von-Neumann statistic assesses the smoothness and stability of the lightcurve by averaging the square differences of consecutive points and dividing by the variance. The least interesting candidates never undergo a significant outburst, leading to a very small value for the Von-Neumann statistic. We exclude sources where the Von-Neumann statistic is less than 0.004.
\section{Results\label{sec:results}}

\begin{figure}[ht]
\begin{center}
\includegraphics[width=0.45\textwidth]{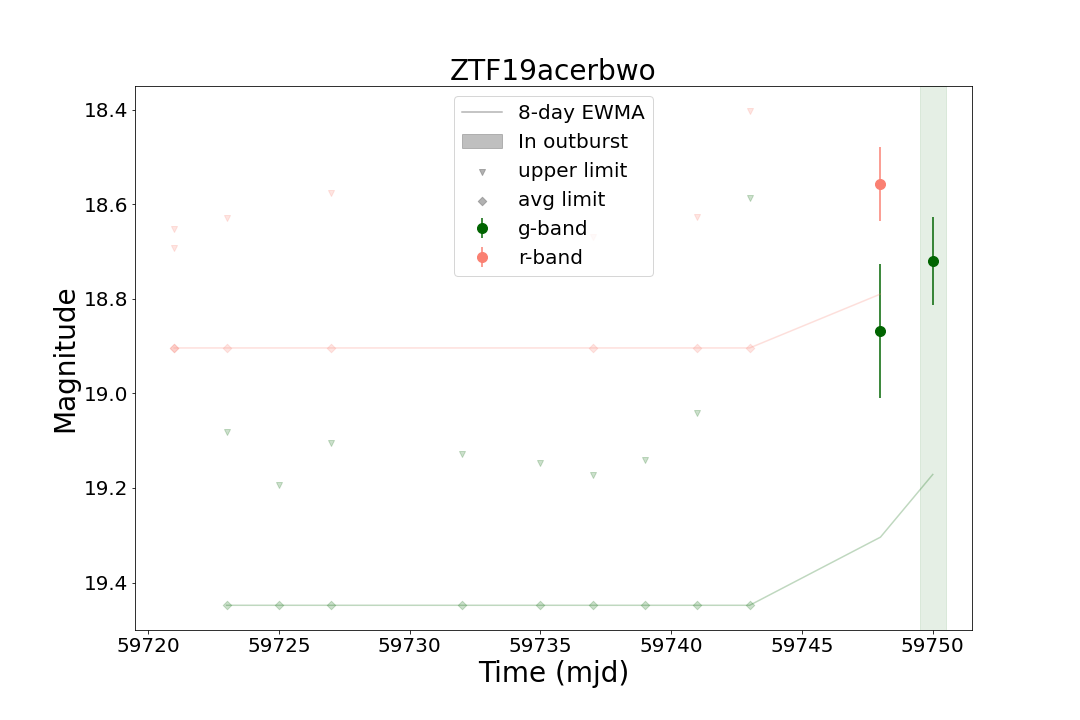}
\includegraphics[width=0.45\textwidth]{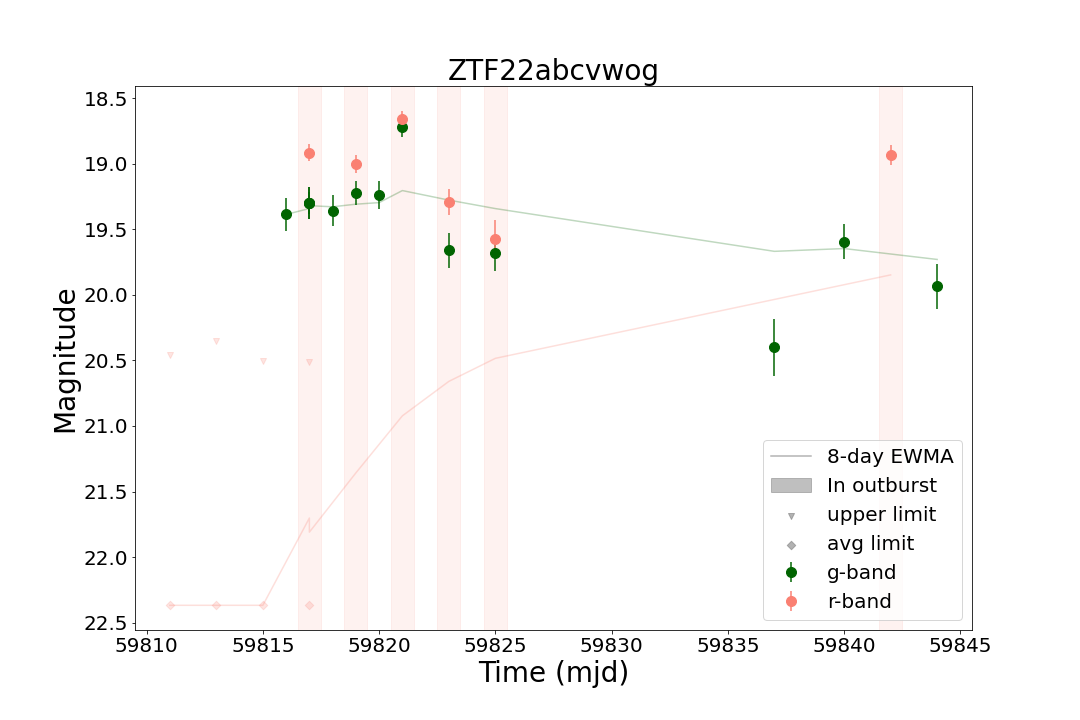}
\caption{Two objects found in our live search. The top plot shows an outburst of MAXI J1957+032, a known LMXB. The bottom plot shows the outburst of Swift J1943.4+0228, an LMXB candidate not observed for 10 years. The red and green colors correspond to observations taken in the ztf-r and ztf-g bands respectively. The lightly plotted lines show the 8-day EMA and the vertical shading indicates observations meeting our outburst metric (these are described by pale grey in the legend).
\label{fig:live_search}}
\end{center}
\end{figure}

\begin{figure}[ht]
\begin{center}
\includegraphics[width=0.45\textwidth]{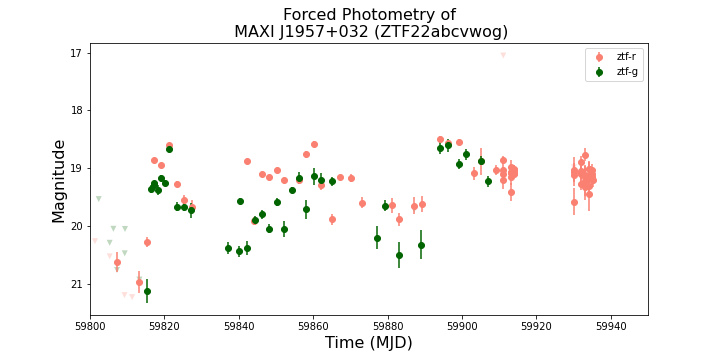}
\caption{ZTF22abcvwog (Swift J1943.4+0228) lightcurve produced by the ZTF forced photometry service showing no previous detections by ZTF, as well as a reflares on MJD=59840 and MJD=59890. 
\label{fig:fps}}
\end{center}
\end{figure}

\begin{figure}[ht]
\begin{center}
\includegraphics[width=0.45\textwidth]{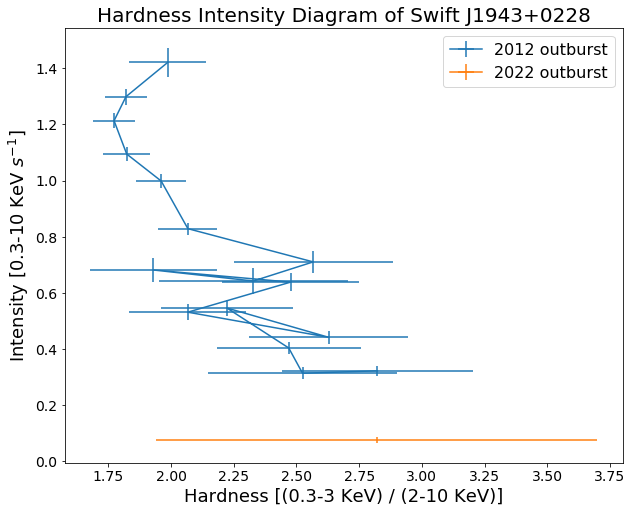}
\caption{Hardness v. intensity of the 2012 and 2022 outbursts of Swift J1943.4+0228. It appears that the 2012 outburst never transitioned into the soft state. The 2022 outburst is barely detected and also harder than the 2012 outburst.
\label{fig:hr}}
\end{center}
\end{figure}

\subsection{Live Search Results}
We identified two LMXB candidates in outburst using our live search pipeline. Figure \ref{fig:live_search} presents the lightcurves and outburst metrics associated with these two sources around the time of discovery. \rev{Table \ref{tab:sources} summarizes these sources along with sources identified in an archival search (\S \ref{sec:archival}).} 

\subsubsection{AMXP MAXI J1957+032}
MAXI J1957+032 is an accreting milisecond puslsar \citep{Sanna:2022:AMXP} that was identified in our live search pipeline as ZTF19acerbwo on June 20th, 2022. This outburst was previously observed and reported by MAXI on June 18th, 2022 \citep{2022ATel15440....1N}. Using the ZTF forced photometry service, we found this source was first observed by ZTF on the same day as the MAXI report \citep{2022ATel15455....1W}. Since the source appeared to be already fading \addtxt{in optical past the followup capabilities of APO}, we did not trigger further followup.

\subsubsection{Swift J1943.4+0228}
Swift J1943.4+0228 is an X-ray transient source first observed in outburst by \textit{Swift} in April 2012 by the Burst Alert Telescope (BAT) \citep{Krimm:2013:Swift_BAT} and subsequently not detected again until it was detected to be in outburst by our pipeline on September 6, 2022 (MJD 59828) as ZTF22abcvwog after having begun an outburst on August 26th (MJD 59817). Ongoing monitoring revealed two more reflares back up to a brightness of $m_r\approx18.5$ beginning on September 18th, 2022 (MJD 59840), each lasting for between 20-40 days. These reflares are visible in the forced photometry lightcurve in figure \ref{fig:fps}. We observed this outburst with the KOSMOS spectrograph \citep{OSMOS:2011} on the 3.5m telescope at APO and were unable to obtain a signal in 40 minutes of observation.

The original \textit{Swift} detection and monitoring revealed a 2 month long decay period in the soft X-rays. We obtained a detection with \textit{Swift} X-ray Telescope (XRT) of around $0.076\,\pm\,0.01\mathrm{\,counts\,s^{-1}}$ count rate, far below the previous 2012 XRT peak detection of $1.4\,\pm\,0.04\mathrm{\,counts\,s^{-1}}$, indicating that this event may be a failed outburst \citep{Alabarta:2021:FailedOutburst}. Additionally, the hardness ratio of our detection was $2.82 \,\pm\,0.88$, harder than the previous outburst, which reached $1.77\,\pm\,0.08$, as shown in the hardness intensity diagram in figure \ref{fig:hr}. 

Using the catalog from \citet{Green:19:3DDustMap}, we estimate the distance to the system to be $1.2\pm 0.3\mathrm{\,kpc}$. Combining this with our measured X-ray flux, we use WebPIMMs and a power law model of the source with photon index of 2 and column density of $1.9\times 10^{21}\mathrm{cm^{-2}}$ \citep{atel:2012:GROND} to estimate the luminosity of the source to be $1.7\times10^{33}\mathrm{\,ergs\,s^{-1}}$. \addtxt{The calculated luminosity of the 2012 ($1.4\times10^{34}\mathrm{\,ergs\,s^{-1}}$) and 2023 outbursts are more characteristic of a cataclysmic variable than an LMXB. However, the long intervals between activity and the X-ray lightcurve of the source, which took over two months to fade appeared similar to a hard-only outburst of an LMXB, making this source worth monitoring for future activity}.

\begin{figure*}[ht]
\begin{center}
\includegraphics[width=0.45\textwidth]{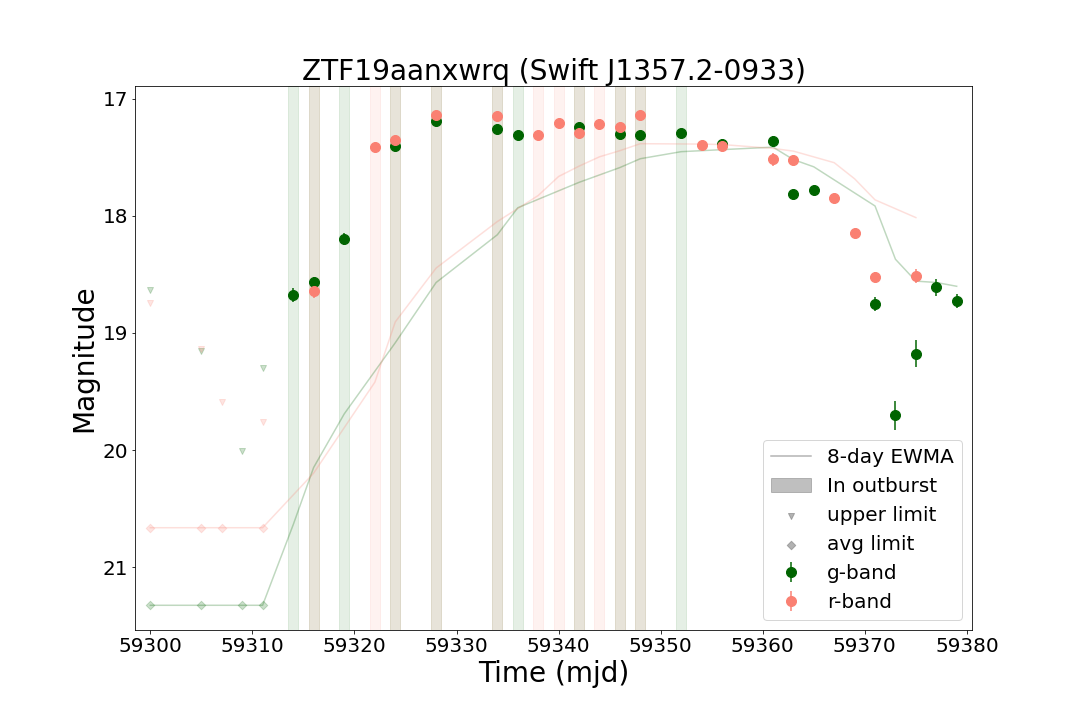}
\includegraphics[width=0.45\textwidth]{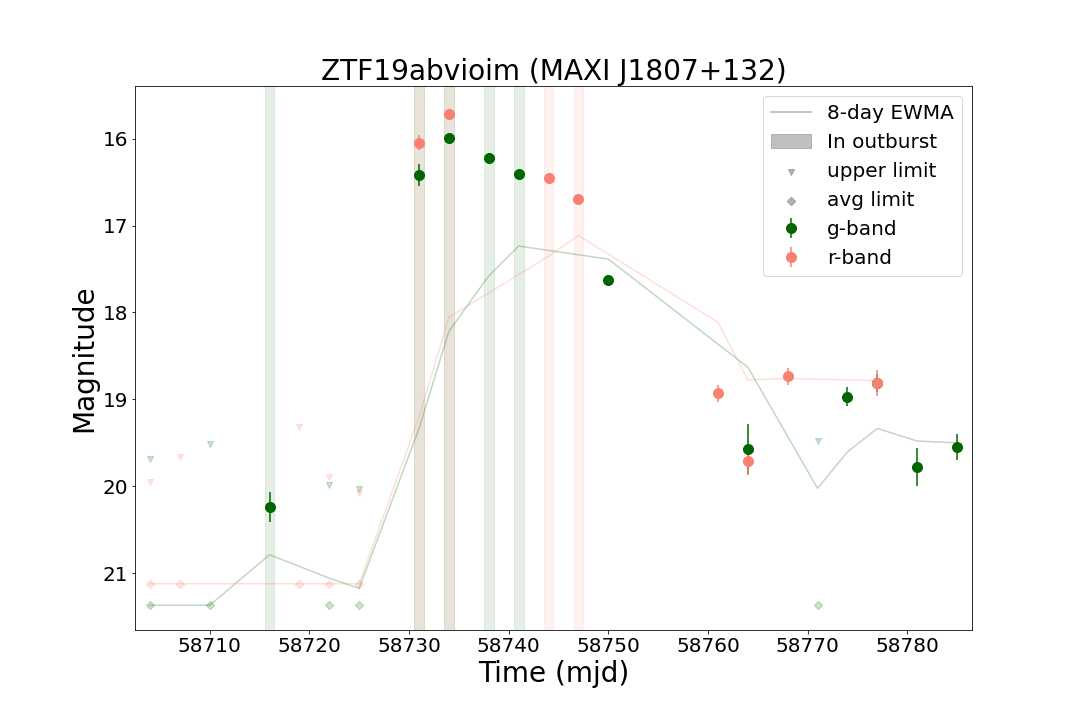}
\includegraphics[width=0.45\textwidth]{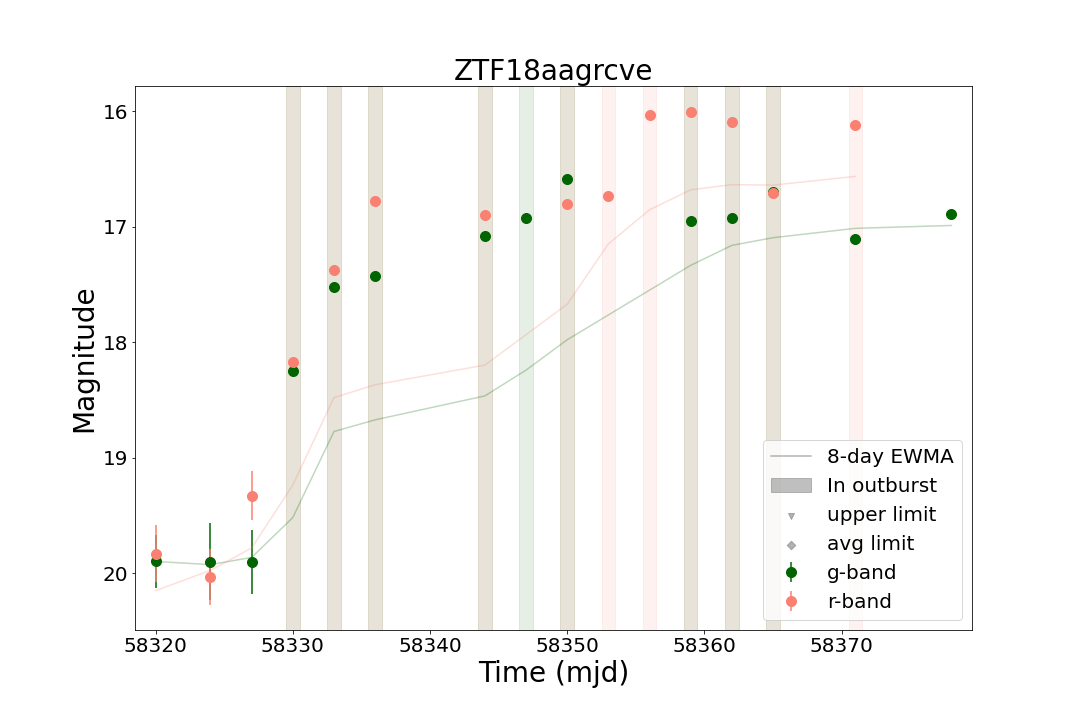}
\includegraphics[width=0.45\textwidth]{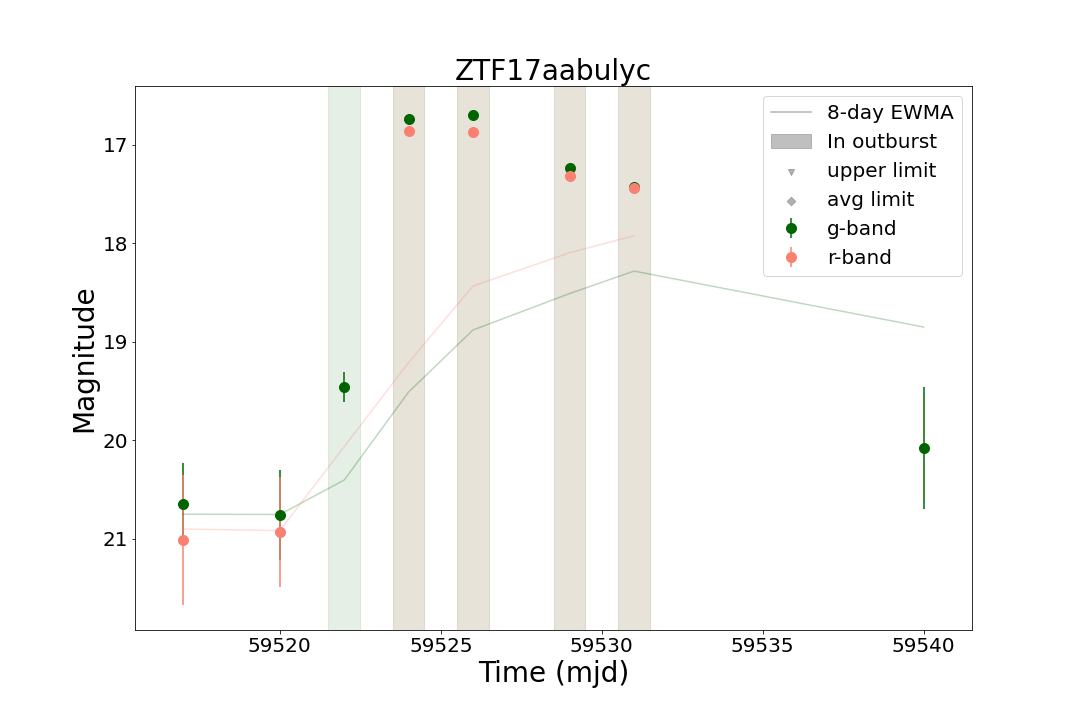}
\caption{Examples of candidates found in the archival search. Both ZTF19aanxwrq and ZTF19abvioim are known LMXBs (Swift J1357.2$-$0933 and MAXI J1807.2$+$132). ZTF18aagrcve displays a rise and sustained outburst profile similar to that of LMXBs; however, a literature search revealed it to be a polar. ZTF17aabulyc is a cataclysmic variable, the short duration of outburst and large duty cycle from its photometric history confirm this.}
\label{fig:archive_search}
\end{center}
\end{figure*}

\subsection{Archival Search Results} \label{sec:archival}
Running the archival search from June 1, 2018 to July 14th, 2022, we crosmatched 75,787 ZTF sources to objects in our combined X-ray catalog. Of these sources, 27952 were left after making the SIMBAD crossmatch cut. After requiring the reference magnitude of the candidate to be 15th mag or fainter and a galactic latitude of less than 15$\deg$, there were 17966 and 7789 sources remaining, respectively.

We then made further cuts on lightcurve features. First, we required at least 4 observations in a single filter, in any 28-day stretch. This narrowed our sample down to 3730 sources. We also required at least 4 observations termed ``in outburst'', with the EMA metric \ref{eq:EMA} greater than 0.25 at least 3 times in a 28-day period, resulting in 2584 candidates remaining. Finally, we required a Von Neumann statistic of at least 0.004 in at least one of the filters, a threshold that we selected after calculating this metric for known X-ray binary outbursts found in the live search, as well as XTE J1859+226 (ZTF21aagyzqr, \citealp{Bellm:2021:ZTF21aagyzqr}) and Swift J1357.2$-$0933 (ZTF19aanxwrq, \citealp{Bellm:2021:ZTF19aanxwrq}). 
Examples of lightcurves passing all cuts are given in figure \ref{fig:archive_search}.

\begin{figure*}[ht]
\begin{center}
\includegraphics[width=0.45\textwidth]{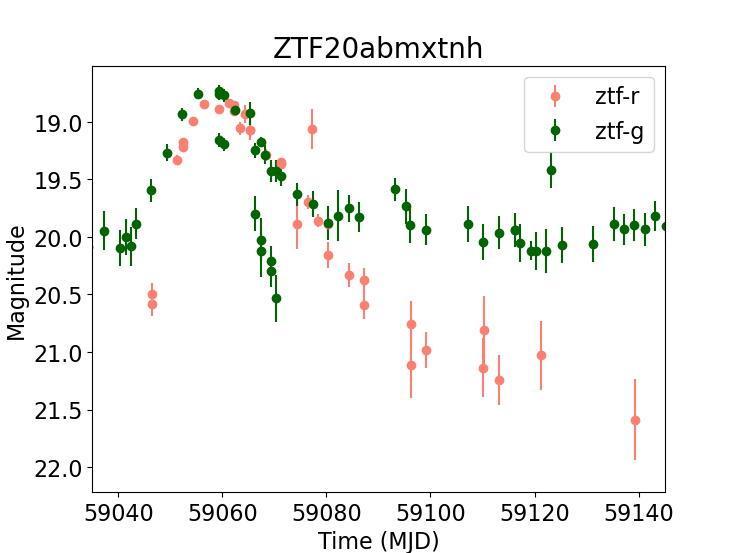}
\includegraphics[width=0.43\textwidth]{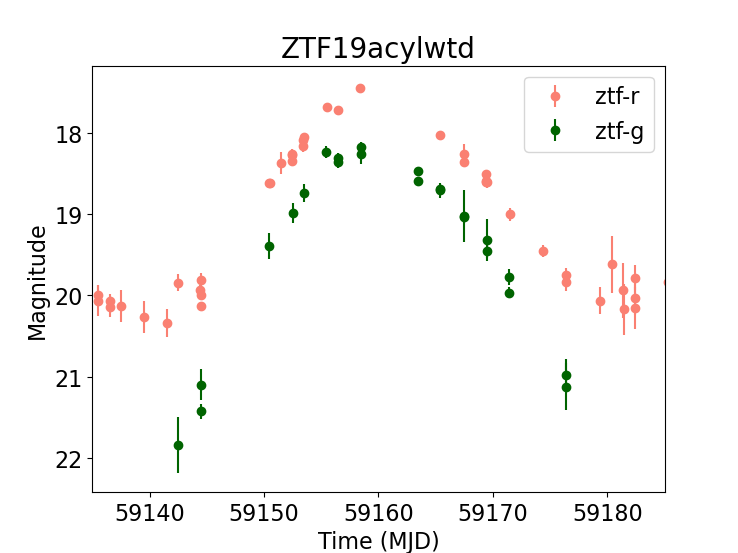}
\includegraphics[width=0.45\textwidth]{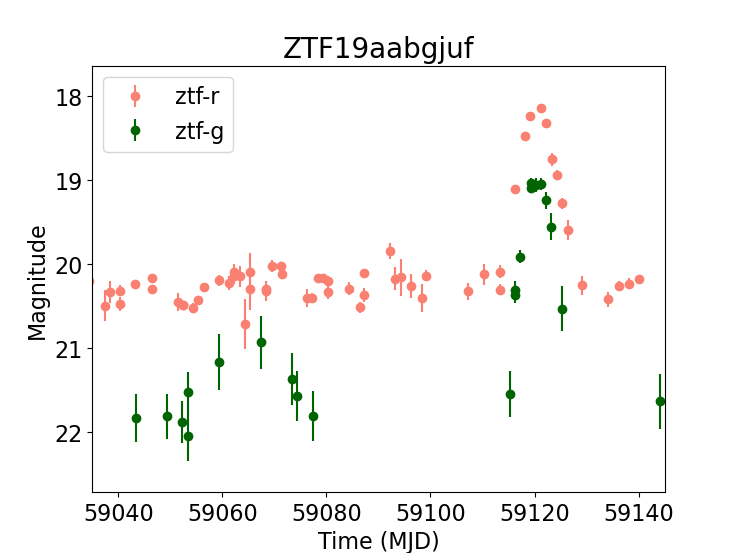}
\caption{3 sources displaying outbursts with quiescent magnitudes below the ZTF alert pipeline detection limit. These lightcurves were obtained using forced photometry and we relaxed the required signal to noise from 5 to 3.}
\label{fig:other_sources}
\end{center}
\end{figure*}

After visual inspections and literature searches for lightcurves displaying characteristics of LMXB outbursting, we recovered \rev{3} known LMXBs in the alert archives:  XTE J1859+226 (ZTF21aagyzqr; \citealp{Bellm:2021:ZTF21aagyzqr}, \citealp{tmp_Bellm:23:XTEJ1859}), Swift J1911.2+0036 (ZTF18accedau), and MAXI J1807+132 \citep[ZTF19abvioim;][]{atel:MAXIJ1807+132}. One source from the live search, Swift J1911.2+0036, was also recovered in the \addtxt{archival} search. We also discovered 3 sources (ZTF20abmxtnh, ZTF19acylwtd, and ZTF19aabgjuf) displaying short outbursts peaking above the limiting magnitude of ZTF; these are shown in figure \ref{fig:other_sources} and discussed below.

Additionally, one known LMXB, Swift J1357.2$-$0933 (ZTF19aanxwrq), was excluded due to the galactic latitude cut.

\subsection{ZTF20abmxtnh}
This source is classified as a Type Ia supernova (SN2020qbw) in TNS \citep{2020qbw:2020:classification}. The location at galactic latitude 9.58$^{\circ}$, as well as the relatively short period of time above the ZTF limiting magnitude enabled it to pass our cuts and to be flagged as a candidate. This source peaked at 18.73 mag in g-band and 18.83 mag in r-band two days later after the first ZTF observation. It was visible to ZTF over a period of roughly 3 months. The lightcurve shows distinct color evolution, with the r-band fading slower than  g-band. The X-ray counterpart for this source was Swift 2SXPS J194138.3+423721. Upon examining the image cutouts, it appears likely that the $g$-band "quiescent" flux is actually contamination from the host galaxy.

\subsection{ZTF19acylwtd}
This source peaked at 17.11 mag in r-band and the total duration of the outburst was around 30 days. The quiescent magnitude of this source was measured by PS1 to be $20.06\,\pm\,0.004$ in the r-band and $21.20\,\pm\,0.05$ in the g band \citep{Panstarrs:2016:PS1}. The X-ray counterpart for this source was Swift 1XPS J070927.3$-$110211. Using a distance of 5 kpc \citep{BailerJones:2021:GaiaeDR3}, we estimate the X-ray luminosity between 0.5 and 7 keV of the source to be $2\times10^{33}\mathrm{\,ergs\,s^{-1}}$. Based on the quiescent magnitude and X-ray luminosity, this is likely a CV.

\subsection{ZTF19aabgjuf}
This source peaked at 17.95 mag in r-band and the total duration of the outburst lasted around 15 days. The quiescent magnitude of this source was measured by PS1 to be $20.26\,\pm\,0.02$ in the r band and $21.60\,\pm\,0.07$ in the g band \citep{Panstarrs:2016:PS1}. Two other potential outbursts with two points in both r and g-band on the rise are seen on the edges of gaps in photometric data. The X-ray counterpart to this object was found by \textit{Chandra}: CSC2 J205921.2+543035. We estimate the X-ray luminosity of this source to be $1 \times 10^{32} \; \mathrm{\,ergs\,s^{-1}}$, assuming a distance of 4 kpc. Based on the quiescent magnitude, recurrence times, and X-ray luminosity, this source is also likely a CV.

\begin{table*}
    \begin{tabular}{|l|l|l|r|r|r|}
        \hline
        ZTF id & Source Name & Source Type & Date Detected (ZTF) & Peak Magnitude\\ 
        \hline
        ZTF19acerbwo & MAXI J1957+032 & LMXB & Jun-20-2022 & 19.1 \\
        ZTF22abcvwog & Swift1943+0228 & LMXB? & Aug-26-2022 & 18.5 \\
        ZTF19abvioim & MAXI J1807+132 & LMXB & Sep-05-2019 & 15.7 \\
        ZTF21aagyzqr & XTE J1859+226 & LMXB & Feb-04-2021 & 18.5 \\
        ZTF18accedau & Aql X-1 & LMXB & May-05-2022 & 16.9 \\
        ZTF19aatqkhi & Swift J0636.6+3536 & CV & Feb-06-2021 & 17.4 \\
        ZTF20abmxtnh & SN2020qbw & SN1a & Jul-16-2020 & 18.7 \\
        ZTF19acylwtd & -- & CV & Oct-28-2020 & 17.5 \\
        ZTF19aabgjuf & -- & CV & Sep-24-2020 & 18.4 \\

        \hline
    \end{tabular}
    \caption{\rev{Sources examined in detail from the live and archival searches.}}
    \label{tab:sources}
\end{table*}

\section{Selection Effects and Estimating the Population of Galactic LMXBs}

We estimated the proportion of expected outbursting systems discoverable by our pipeline by simulating a population of Galactic LMXBs with a bulge/disk/halo distribution \citep{Grimm:02:XRBDistribution} and subjecting the simulated population to a selection function approximating the cuts of our search. Each simulated object is assigned parameters such as recurrence time, mean outburst duration, and duty cycle drawn from 45 sources from \addtxt{table 8 of }the WATCHDOG catalog from \cite{Tet16} of transient black hole X-ray binaries.
\footnote{We excluded the two long term transient sources \addtxt{(SWIFTJ1753.5$-$0127 and GRS1915+105) and two sources with high mass companions (SAXJ1819.3$-$2525 and XTEJ0421+560).}} 
Using bootstrapping, we obtain a maximum likelihood estimate of the number of galactic LMXBs based on our observation of \rev{4} outbursting objects. We do this by subsampling a number N of simulated objects, and passing the subsample through a selection function that replicates our pipeline, including limits due to the survey depth and area of ZTF, the exclusion of ZTF alerts without an X-ray counterpart, and the likelihood of an outburst during our 4 year search window. 

\subsection{ZTF Survey Cut}
We used ZTF pointing data to determine the number of simulated objects within ZTF fields (38.9\%) that were observed more than 100 times (37.8\%). We calculated the extinction to each simulated position using three-dimensional dust maps \citep{Green:19:3DDustMap,Lallement:19:3DDust}, assumed a \addtxt{fiducial outburst brightness increase of 6 magnitudes based on BlackCAT outburst magnitudes and distances},
and used the simulated distances \addtxt{and absolute magnitudes} to calculate the apparent magnitudes of each source. \addtxt{Absolute magnitudes were simulated using the orbital period-magnitude relation from \cite{Casares:2018:PhotometricMassFunction}, and by sampling known orbital periods and perturbing them by up to 50\%.} Using a fiducial limiting magnitude of m = 20.5 for ZTF, we find that around 75\% of outbursting systems reach a brightness greater than 20.5 mag. Additionally, we note that the galactic latitude cut ($|b| \leq 15\deg$) in the archival search excluded less than 4\% of the simulated sources, \blue{although} \addtxt{our model for the spatial distribution of LMXBs \citep{Grimm:02:XRBDistribution} may underestimate the number of LMXBs in the halo,  due to natal kicks \citep{Gandhi:2020:LMXBspatial}.}

\begin{figure}
    \centering
    \includegraphics[width=0.45\textwidth]{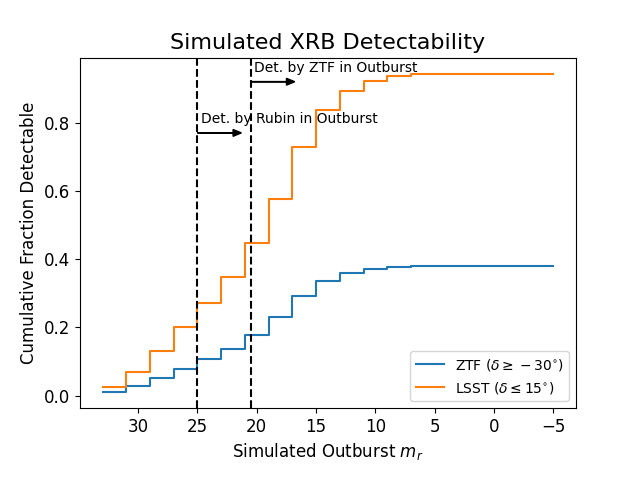}
    \caption{Cumulative fraction of simulated LMXBs discoverable by the ZTF and LSST surveys accounting for distance to each object, extinction using 3D dust maps \citep{Green:19:3DDustMap,Lallement:19:3DDust}, and location of the observatory. The single-image limiting magnitudes for ZTF and for Rubin are plotted as vertical dashed lines. Almost all simulated outbursting XRBs are observable by Rubin Observatory, while only about 39\% are expected to be in the ZTF survey footprint.}
    \label{fig:cumldisc}
\end{figure}

\subsection{Cut on Sources with X-ray Counterparts}
Our decision to require a crossmatch between each alert to a catalog of X-ray sources allows us to investigate a more tractable number of candidates. However, the aggressiveness of this cut means that we may be excluding many sources that have never been detected by any X-ray survey. To obtain an order of magnitude estimate of the number of LMXBs potentially eliminated by this cut, we assumed a quiescent X-ray luminosity of $10^{31}$\,ergs\,s$^{-1}$ and an outburst luminosity of $10^{38}$\,ergs\,s$^{-1}$ and calculated the X-ray flux from each source in quiescence and outburst, using the simulated distances. We then used WebPIMMs to estimate the counts per second for each simulated source at quiescence and in outburst, while accounting for extinction. Finally, we estimated the percentage of sources expected at a signal to noise ratio of 5 for each of the catalogs in our crossmatch (table \ref{tab:cat}) by sampling the exposure times in each catalog and multiplying by the extinction-corrected flux to estimate the total counts detected per simulated source. We use $\eta$ to represent the percentage of sources expected, for each of the 5 catalogs, for sources in both quiescent and outburst states. 

While sources in outburst are almost always bright enough to be observed by each instrument, most sources are in the outburst state only a small fraction of time. This fraction, known as the duty cycle, is observed to be anywhere between less than 1\% to over 50\% \citep{Tet16}, with a median in the WATCHDOG database of black hole XRBs of 2.7\% and a mean of 10\%. We use Monte Carlo sampling to simulate whether each system is in outburst or quiescence and then use Monte Carlo sampling again to simulate whether the system is detected and cataloged, where $\eta$ describes the probability of discovery. This sampling procedure is repeated for each of the 5 catalogs. A simulated source is tagged as passing this cut if it appears in any one of the catalogs. Using this method, we estimate that around 8.8\% of galactic LMXBs are present in our unified X-ray catalog (see table \ref{tab:pct_det}).

By using sampled parameters from the WATCHDOG database, we are assuming that the outburst parameters (mean outburst duration, duty cycle, and recurrence time) of the known sources are representative of undiscovered LMXBs, including systems with neutron stars. This is unlikely because of observational biases--sources that are in outburst more often and more recently are more likely to be discovered. Additionally, many of the WATCHDOG sources have only one recorded outburst, meaning that the duty cycle and recurrence times are upper limits. To account for this, we also run this step of the selection function using the median duty cycle, which could be more representative of the galactic population. Using this method, we estimate that around 4.4\% of galactic LMXBs are present in our unified X-ray catalog.

\subsection{Outbursting LMXBs}
We then estimate the proportion of LMXBs that undergo an outburst in any 4 year period using the WATCHDOG catalog of known black hole XRBs \citep{Tet16}, calculating the number of expected outbursts for each simulated object detectable in a $T=4\,$year period using the recurrence time:
\begin{equation} \label{eq:exp_ob}
    Expected(N_{outbursts}) = \frac{T + t_{outburst}}{t_{recurrence}}
\end{equation}
where $T$ is the length of the archival search (4 years). Since many of the WATCHDOG sources have recurrence times longer than 4 years, only a fraction of the sources will undergo a state change in our data. We can then calculate the probability of at least one outburst using the cumulative distribution function of the Poisson distribution with $\lambda = \mathbb{E}(N_{outbursts})$:
\begin{equation}
    P_{outburst} = 1 - P(0)
\end{equation}
Again using Monte Carlo sampling, we can simulate the sources that undergo at least one outburst during the 4 year period. 

We assume the X-ray outburst parameters are representative of the optical outbursts. However, there is evidence that short optical outbursts can happen without a state change or full X-ray brightening, such as in misfired or failed outbursts (e.g., \citealp{Alabarta:2021:FailedOutburst, Baglio:2022:CenX4_failed_ob, tmp_Bellm:23:XTEJ1859}) and possibly the outburst of Swift J1943+0228 reported in this paper. As a result our cut here may be too stringent, as systems may undergo optical outburst more frequently, particularly as activity without the substantial X-ray brightening needed to trigger X-ray monitors.

\subsection{Estimate of the Galactic Population}
Finally, we performed a Monte Carlo simulation to estimate the number of interacting LMXBs in the galaxy consistent with our observations. We do this by taking 10,000 samples of size N from our simulated dataset, passing them through the cuts described above, and finding the percentage of samples returning \blue{4} sources. 

Using the sampled outburst parameters, we estimate that \addtxt{1.6\%} of sources in outburst are detectable. Accounting for the duration of our survey, we find that we detect \addtxt{0.76\%} of the total population of LMXBs and obtain an estimate of the number of sources with a 95\% confidence interval (CI) of \addtxt{$530^{+810}_{-320}$}. This estimate is lower than current estimates of the number of LMXBs in the galaxy. The 2016 BlackCAT survey \citep{Corral-Santana:15:BlackCAT} estimates a total of 1300 galactic BHXBs, implying around 4000 LMXBs with either black hole or neutron star primaries, assuming that there are roughly double the number of neutron star LMXBs as black hole XRBs \citep{Bahramian:2022:LMXBReview}.

If we substitute the median duty cycle instead of the sampling this parameter, we estimate that \addtxt{0.79\% of outbursts are detectable, meaning we detect 0.012\% of the galactic population, giving to a total of $3390^{+3980}_{-1930}$} galactic LMXBs. This appears to be more in line with current estimates. However, because we have only \rev{4} confirmed objects recovered by the pipeline, our results are accompanied by large uncertainties. The large difference between the two estimates is also due to the frequency of expected outbursts, with many fewer outbursts expected for an undiscovered population with the fiducial duty cycle.

\subsection{\addtxt{Discovery of Outbursts in Optical vs. X-ray}}

\addtxt{While accretion outbursts of LMXB are intrinsically much energetic at X-ray wavelengths, the greater single epoch sensitivities of optical observatories means that the flux from many outbursting LMXBs first exceed the sensitivity limit of an optical survey than an all sky X-ray monitor. Using 28 black hole XRBs from the BlackCAT catalog \citep{Corral-Santana:2016:BlackCAT} with peak X-ray flux counts, optical or IR (OIR) peak outburst magnitude, and distance estimates, we calculated both the X-ray and OIR luminosities, correcting for extinction. We then fit a 2-component Gaussian mixture model to the resulting set of luminosities and sampled this model additional X-ray and OIR luminosities, which we combined with distances and extinctions to obtain a simulated set of outburst fluxes in both X-ray and OIR.}

\addtxt{Taking a ZTF limiting AB magnitude of 20.5 and a single orbit MAXI sensitivity of 10 milliCrab between 2--30\,keV, we calculated the relative sensitivities of these observatories to be $3.3\times10^{-14}\mathrm{\,ergs\,s^{-1}\,cm^{-2}}$ and $1.21\times10^{-10}\mathrm{\,ergs\,s^{-1}\,cm^{-2}}$. We assume that if the OIR flux to X-ray flux ratio is greater than the MAXI to ZTF flux ratio, than the source will first exceed the sensitivity of ZTF before MAXI. Using this metric, we estimate 21\% of outbursts would first be visible in the optical. This estimate is likely conservative: 
we assume that the ratio of optical flux to x-ray flux at the peak of the outburst is similar to the ratio during the rise, while models for accretion disk outbursts may suggest that the optical radiation begins to increase before the X-ray increase  \citep{Rus19, Ham:2020:DIMReview}. The Einstein Probe is currently slated for launch by the end of 2023 and carries an ASM with an order of magnitude sensitivity increase compared to current ASMs \rev{ \citep{Chen:2022:EP}}. Comparing this upcoming probe to LSST, and assuming a sensitivity 10x that of MAXI, we calculate that about 27\% of of outbursts would first be detectable to LSST in the optical. }

\addtxt{Additionally we ran ZTF forced photometry on all outbursting LMXBs in the past 5 years with a known optical counterpart. Out of the 11 LMXBs with ZTF observations, we find that 5 of the sources (Swift J1858.6$-$0814, Swift J1357.2$-$0933, Aquila X-1, XTE J1859+226,  MAXI J1807+132; see Figure \ref{fig:optfirst}) had at least one detection by ZTF before the time of first detection in X-ray, and that 3 of these sources (MAXI J1820+070, Swift J1910.2$-$0546, MAXI J1957+032; Figure \ref{fig:optxray}) showed concurrent optical and x-ray outbursts.
Three sources (4U 1730$-$22, MAXI J1816$-$195, IGR J17591$-$2342; Figure \ref{fig:xrayonly}) showed no visible increase in optical brightness.
}

\subsection{Optical Discovery of LMXB Outbursts in the Rubin Era}
As shown in figure \ref{fig:cumldisc}, due to the location of Palomar in the Northern Hemisphere as well as the moderate limiting magnitude of ZTF, we only expect $<40$\% of outbursting LMXBs to be present in ZTF data. We can estimate the expected proportion of outbursting systems discoverable by Rubin's LSST using survey cadence simulations \citep{Bianco:22:CadenceOptimization, PSTN-055}, plus our simulation of XRB positions following \citet{Johnson:19:LSSTLMXB} and \citet{Grimm:02:XRBDistribution}. The location of Rubin Observatory in the Southern Hemisphere means that it will be much better positioned to observe the Galactic Bulge, and will observe approximately 86\% of the outbursts. Using an XRB metric\footnote{\url{https://github.com/lsst/rubin_sim/blob/main/rubin_sim/maf/maf_contrib/xrb_metrics.py}} and v3.0 of the LSST baseline observing cadence \citep{PSTN-055}, we simulated 10,000 outbursts following a fast rise exponential decay template \citep{Chen:1997} and estimate that 43.3\% of outbursts will be detectable in 10-year survey time of LSST by Rubin.

Much of the improvement (in the previous section we estimated our search using ZTF to detect 2.4\% and 1.2\% of all outbursts, depending on the assumptions of the distribution of the duty cycle of undiscovered objects) is due to the elimination of the the X-ray counterpart cut, which eliminates between 91-96\% of LMXBs.  Using equation \ref{eq:exp_ob} we estimate 26\% of systems to undergo at least one outburst in a ten year period. As a result, we expect LSST to discover outbursts from approximately 10\% of all galactic LMXBs, i.e. roughly 400 objects.

\addtxt{Notably, most of the outbursts we observed and recovered in the live and archival searches were shorter (on order weeks) than typical outbursts which can last for months to years. The duration of our search of roughly 4 years means that we will preferentially observe sources with shorter duty cycles, which tend to correlate with shorter outburst duration.}

\begin{table*}
    \begin{tabular}{|l|l|l|r|r|r|}
        \hline
        Observatory & Catalog & Sky Coverage& $\eta_{quiescent}$ & $\eta_{outburst}$\\ 
        \hline
        \rosat & 2RXS & 100\% & 0.0025\% & 99.999\%  \\
        \xmm & XMMSL2 & 84\% & 0.013\% & 37.4\% \\
        \swift-XRT & 2SXPS & 9.1\% & 0.14\% & 100\% \\ 
        \xmm & 4XMM-DR10 & 3.0\%  & 17.8\% & 100\% \\ 
        \chandra & CSC 2.0 & 1.3\% & 1.6\% & 99.2\% \\
        \hline
    \end{tabular}
    \caption{Estimate of percentage of galactic LMXBs that are expected ($\eta$) in each of the major catalogs we crossmatch to.}
    \label{tab:pct_det}
\end{table*}

\section{Discussion\label{sec:discussion}}
The previous section demonstrates that requiring a candidate to have a crossmatched X-ray source is a fairly stringent criteria. Our order of magnitude estimate of the percentage of LMXBs catalogued by one of the 5 X-ray surveys we crossmatch against is 4.4\% to 8.9\%, meaning our pipeline excludes more than 9 out of 10 potential candidates. One example of an LMXB candidate missed by both our live and archival searches is ZTF19acwrvzk \citep[AT2019wey;][]{Yao21}, which did not have a known X-ray counterpart at the time of detection. We also have implemented a pipeline without a crossmatching criteria, instead using EMAs of different exponential decay timescales to detect outbursting events, as well as additional timeseries features to make cuts on both the live alert stream, and the archival candidates.

The presence of only very faint hard X-ray emission associated with the optical flaring and reflaring of Swift J1943+0228 may be indicative of repeated failure of the heating front to propagate throughout the disk. If Swift J1943+0228 is indeed an LMXB, the previously detected 2012 outburst appears also to have never transitioned into the soft state. \citet{Tet16} found that around 39\% of the outbursts present in the WATCHDOG catalog did not transition into the soft state and that these hard-only outbursts represent a substantial component of all events. Postulating that these hard-only outbursts represent sources that did not reach the required mass transfer rate to transition to the soft state, it is possible that these optical outbursts represent the lowest energy tails of these phenomena, where the outburst stalls before increased mass transfer leads to brightening in the hard state. 

Because almost all XRB outbursts, even failed transition outbursts, are discovered due to X-ray brightening, untargeted searches at optical wavelengths of outbursts can provide a powerful complementary pathway to the discovery and characterization of LMXBs and their outbursts \citep[e.g.,][]{tmp_Bellm:23:XTEJ1859}. Due to their increased sensitivity to low luminosity and earlier activity, optical searches will offer a unique window into a population of outbursts that are not well understood. While we only have a few examples of this behavior in the ZTF data, the location and greater depth of LSST will enable observations of a greater number of LMXB outbursts, including observations earlier, closer to outburst onset, and allow us to build a robust sample of optical outbursts.

\section*{Acknowledgement}
We thank Roy Williams for suggesting use of the exponentially weighted moving average to identify outbursting sources.

Based on observations obtained with the Samuel Oschin Telescope 48-inch and the 60-inch Telescope at the Palomar Observatory as part of the Zwicky Transient Facility project. ZTF is supported by the National Science Foundation under Grants No. AST-1440341 and AST-2034437 and a collaboration including current partners Caltech, IPAC, the Weizmann Institute of Science, the Oskar Klein Center at Stockholm University, the University of Maryland, Deutsches Elektronen-Synchrotron and Humboldt University, the TANGO Consortium of Taiwan, the University of Wisconsin at Milwaukee, Trinity College Dublin, Lawrence Livermore National Laboratories, IN2P3, University of Warwick, Ruhr University Bochum, Northwestern University and former partners the University of Washington, Los Alamos National Laboratories, and Lawrence Berkeley National Laboratories. Operations are conducted by COO, IPAC, and UW.

The ZTF forced-photometry service was funded under the Heising-Simons Foundation grant \#12540303 (PI: Graham).

YK and ECB gratefully acknowledge support from the NSF AAG grant 1812779 and grant \#2018-0908 from the Heising-Simons Foundation.

ECB acknowledges further support from the Vera C.\ Rubin Observatory, which is supported in part by the National Science Foundation through
Cooperative Agreement 1258333 managed by the Association of Universities for Research in Astronomy
(AURA), and the Department of Energy under Contract No. DE-AC02-76SF00515 with the SLAC National
Accelerator Laboratory. Additional LSST funding comes from private donations, grants to universities,
and in-kind support from LSSTC Institutional Members.

AM acknowledges financial support from Padova University, Department of Physics and Astronomy Research Project 2021 (PRD 2021)

\bibliographystyle{aasjournal}
\bibliography{main}

\begin{thebibliography}{}
\expandafter\ifx\csname natexlab\endcsname\relax\def\natexlab#1{#1}\fi
\providecommand{\url}[1]{\href{#1}{#1}}
\providecommand{\dodoi}[1]{doi:~\href{http://doi.org/#1}{\nolinkurl{#1}}}
\providecommand{\doeprint}[1]{\href{http://ascl.net/#1}{\nolinkurl{http://ascl.net/#1}}}
\providecommand{\doarXiv}[1]{\href{https://arxiv.org/abs/#1}{\nolinkurl{https://arxiv.org/abs/#1}}}

\bibitem[{{Alabarta} {et~al.}(2021){Alabarta}, {Altamirano}, {M{\'e}ndez},
  {C{\'u}neo}, {Vincentelli}, {Castro-Segura}, {Garc{\'\i}a}, {Luff}, \&
  {Veledina}}]{Alabarta:2021:FailedOutburst}
{Alabarta}, K., {Altamirano}, D., {M{\'e}ndez}, M., {et~al.} 2021, \mnras, 507,
  5507, \dodoi{10.1093/mnras/stab2241}

\bibitem[{{Armas Padilla} {et~al.}(2014){Armas Padilla}, {Wijnands},
  {Degenaar}, {Mu{\~n}oz-Darias}, {Casares}, \&
  {Fender}}]{Padilla:2014:FaintestBH}
{Armas Padilla}, M., {Wijnands}, R., {Degenaar}, N., {et~al.} 2014, \mnras,
  444, 902, \dodoi{10.1093/mnras/stu1487}

\bibitem[{Baglio {et~al.}(2022)Baglio, Saikia, Russell, Homan, Waterval,
  Bramich, Campana, Lewis, den Eijnden, Alabarta, Covino, D’Avanzo, Goldoni,
  Masetti, \& Muñoz-Darias}]{Baglio:2022:CenX4_failed_ob}
Baglio, M.~C., Saikia, P., Russell, D.~M., {et~al.} 2022, The Astrophysical
  Journal, 930, 20, \dodoi{10.3847/1538-4357/ac63ad}

\bibitem[{{Bahramian} \& {Degenaar}(2022)}]{Bahramian:2022:LMXBReview}
{Bahramian}, A., \& {Degenaar}, N. 2022, arXiv e-prints, arXiv:2206.10053,
  \dodoi{10.48550/arXiv.2206.10053}

\bibitem[{{Bailer-Jones} {et~al.}(2021){Bailer-Jones}, {Rybizki}, {Fouesneau},
  {Demleitner}, \& {Andrae}}]{BailerJones:2021:GaiaeDR3}
{Bailer-Jones}, C.~A.~L., {Rybizki}, J., {Fouesneau}, M., {Demleitner}, M., \&
  {Andrae}, R. 2021, \aj, 161, 147, \dodoi{10.3847/1538-3881/abd806}

\bibitem[{{Balbus} \& {Hawley}(1998)}]{Balbus:1998:InstabilityInACs}
{Balbus}, S.~A., \& {Hawley}, J.~F. 1998, Reviews of Modern Physics, 70, 1,
  \dodoi{10.1103/RevModPhys.70.1}

\bibitem[{{Bellm}(2021{\natexlab{a}})}]{Bellm:2021:ZTF21aagyzqr}
{Bellm}, E.~C. 2021{\natexlab{a}}, The Astronomer's Telegram, 14372, 1

\bibitem[{{Bellm}(2021{\natexlab{b}})}]{Bellm:21:SwiftJ1357ATel}
---. 2021{\natexlab{b}}, The Astronomer's Telegram, 14539, 1

\bibitem[{{Bellm}(2021{\natexlab{c}})}]{Bellm:2021:ZTF19aanxwrq}
---. 2021{\natexlab{c}}, The Astronomer's Telegram, 14539, 1

\bibitem[{{Bellm} {et~al.}(2019{\natexlab{a}}){Bellm}, {Kulkarni}, {Graham},
  {Dekany}, {Smith}, {Riddle}, {Masci}, {Helou}, {Prince}, {Adams},
  {Barbarino}, {Barlow}, {Bauer}, {Beck}, {Belicki}, {Biswas}, {Blagorodnova},
  {Bodewits}, {Bolin}, {Brinnel}, {Brooke}, {Bue}, {Bulla}, {Burruss}, {Cenko},
  {Chang}, {Connolly}, {Coughlin}, {Cromer}, {Cunningham}, {De}, {Delacroix},
  {Desai}, {Duev}, {Eadie}, {Farnham}, {Feeney}, {Feindt}, {Flynn},
  {Franckowiak}, {Frederick}, {Fremling}, {Gal-Yam}, {Gezari}, {Giomi},
  {Goldstein}, {Golkhou}, {Goobar}, {Groom}, {Hacopians}, {Hale}, {Henning},
  {Ho}, {Hover}, {Howell}, {Hung}, {Huppenkothen}, {Imel}, {Ip}, {Ivezi{\'c}},
  {Jackson}, {Jones}, {Juric}, {Kasliwal}, {Kaspi}, {Kaye}, {Kelley},
  {Kowalski}, {Kramer}, {Kupfer}, {Landry}, {Laher}, {Lee}, {Lin}, {Lin},
  {Lunnan}, {Giomi}, {Mahabal}, {Mao}, {Miller}, {Monkewitz}, {Murphy},
  {Ngeow}, {Nordin}, {Nugent}, {Ofek}, {Patterson}, {Penprase}, {Porter},
  {Rauch}, {Rebbapragada}, {Reiley}, {Rigault}, {Rodriguez}, {van Roestel},
  {Rusholme}, {van Santen}, {Schulze}, {Shupe}, {Singer}, {Soumagnac}, {Stein},
  {Surace}, {Sollerman}, {Szkody}, {Taddia}, {Terek}, {Van Sistine}, {van
  Velzen}, {Vestrand}, {Walters}, {Ward}, {Ye}, {Yu}, {Yan}, \&
  {Zolkower}}]{Bellm:2019:ZTFsystem}
{Bellm}, E.~C., {Kulkarni}, S.~R., {Graham}, M.~J., {et~al.}
  2019{\natexlab{a}}, \pasp, 131, 018002, \dodoi{10.1088/1538-3873/aaecbe}

\bibitem[{{Bellm} {et~al.}(2019{\natexlab{b}}){Bellm}, {Kulkarni}, {Barlow},
  {Feindt}, {Graham}, {Goobar}, {Kupfer}, {Ngeow}, {Nugent}, {Ofek}, {Prince},
  {Riddle}, {Walters}, \& {Ye}}]{Bellm:2019:Scheduler}
{Bellm}, E.~C., {Kulkarni}, S.~R., {Barlow}, T., {et~al.} 2019{\natexlab{b}},
  \pasp, 131, 068003, \dodoi{10.1088/1538-3873/ab0c2a}

\bibitem[{{Bellm} {et~al.}(2023){Bellm}, {Wang}, {van Roestel}, {Phillipson},
  {Coughlin}, {Tomsick}, {Groom}, {Healy}, {Purdum}, {Rusholme}, {Sollerman},
  {Bealo}, {Lora}, {Muyllaert}, {Peretto}, \&
  {Schwendeman}}]{tmp_Bellm:23:XTEJ1859}
{Bellm}, E.~C., {Wang}, Y., {van Roestel}, J., {et~al.} 2023, arXiv e-prints,
  arXiv:2309.10742, \dodoi{10.48550/arXiv.2309.10742}

\bibitem[{{Bianco} {et~al.}(2022){Bianco}, {Ivezi{\'c}}, {Jones}, {Graham},
  {Marshall}, {Saha}, {Strauss}, {Yoachim}, {Ribeiro}, {Anguita}, {Bauer},
  {Bauer}, {Bellm}, {Blum}, {Brandt}, {Brough}, {Catelan}, {Clarkson},
  {Connolly}, {Gawiser}, {Gizis}, {Hlo{\v{z}}ek}, {Kaviraj}, {Liu}, {Lochner},
  {Mahabal}, {Mandelbaum}, {McGehee}, {Neilsen}, {Olsen}, {Peiris}, {Rhodes},
  {Richards}, {Ridgway}, {Schwamb}, {Scolnic}, {Shemmer}, {Slater}, {Slosar},
  {Smartt}, {Strader}, {Street}, {Trilling}, {Verma}, {Vivas}, {Wechsler}, \&
  {Willman}}]{Bianco:22:CadenceOptimization}
{Bianco}, F.~B., {Ivezi{\'c}}, {\v{Z}}., {Jones}, R.~L., {et~al.} 2022, \apjs,
  258, 1, \dodoi{10.3847/1538-4365/ac3e72}

\bibitem[{{Boller} {et~al.}(2016){Boller}, {Freyberg}, {Tr{\"u}mper}, {Haberl},
  {Voges}, \& {Nandra}}]{Boller:16:2RXS}
{Boller}, T., {Freyberg}, M.~J., {Tr{\"u}mper}, J., {et~al.} 2016, \aap, 588,
  A103

\bibitem[{{Casares}(2016)}]{Casares:2016:MassRatioHa}
{Casares}, J. 2016, \apj, 822, 99, \dodoi{10.3847/0004-637X/822/2/99}

\bibitem[{{Casares}(2018)}]{Casares:2018:PhotometricMassFunction}
---. 2018, \mnras, 473, 5195, \dodoi{10.1093/mnras/stx2690}

\bibitem[{{Casares} \& {Torres}(2018)}]{Casares:2018:PhotXRBs}
{Casares}, J., \& {Torres}, M. A.~P. 2018, \mnras, 481, 4372,
  \dodoi{10.1093/mnras/sty2570}

\bibitem[{{Chambers} {et~al.}(2016){Chambers}, {Magnier}, {Metcalfe},
  {Flewelling}, {Huber}, {Waters}, {Denneau}, {Draper}, {Farrow}, {Finkbeiner},
  {Holmberg}, {Koppenhoefer}, {Price}, {Rest}, {Saglia}, {Schlafly}, {Smartt},
  {Sweeney}, {Wainscoat}, {Burgett}, {Chastel}, {Grav}, {Heasley}, {Hodapp},
  {Jedicke}, {Kaiser}, {Kudritzki}, {Luppino}, {Lupton}, {Monet}, {Morgan},
  {Onaka}, {Shiao}, {Stubbs}, {Tonry}, {White}, {Ba{\~n}ados}, {Bell},
  {Bender}, {Bernard}, {Boegner}, {Boffi}, {Botticella}, {Calamida},
  {Casertano}, {Chen}, {Chen}, {Cole}, {Deacon}, {Frenk}, {Fitzsimmons},
  {Gezari}, {Gibbs}, {Goessl}, {Goggia}, {Gourgue}, {Goldman}, {Grant},
  {Grebel}, {Hambly}, {Hasinger}, {Heavens}, {Heckman}, {Henderson}, {Henning},
  {Holman}, {Hopp}, {Ip}, {Isani}, {Jackson}, {Keyes}, {Koekemoer}, {Kotak},
  {Le}, {Liska}, {Long}, {Lucey}, {Liu}, {Martin}, {Masci}, {McLean}, {Mindel},
  {Misra}, {Morganson}, {Murphy}, {Obaika}, {Narayan}, {Nieto-Santisteban},
  {Norberg}, {Peacock}, {Pier}, {Postman}, {Primak}, {Rae}, {Rai}, {Riess},
  {Riffeser}, {Rix}, {R{\"o}ser}, {Russel}, {Rutz}, {Schilbach}, {Schultz},
  {Scolnic}, {Strolger}, {Szalay}, {Seitz}, {Small}, {Smith}, {Soderblom},
  {Taylor}, {Thomson}, {Taylor}, {Thakar}, {Thiel}, {Thilker}, {Unger},
  {Urata}, {Valenti}, {Wagner}, {Walder}, {Walter}, {Watters}, {Werner},
  {Wood-Vasey}, \& {Wyse}}]{Panstarrs:2016:PS1}
{Chambers}, K.~C., {Magnier}, E.~A., {Metcalfe}, N., {et~al.} 2016, arXiv
  e-prints, arXiv:1612.05560, \dodoi{10.48550/arXiv.1612.05560}

\bibitem[{{Chen} {et~al.}(1997){Chen}, {Shrader}, \& {Livio}}]{Chen:1997}
{Chen}, W., {Shrader}, C.~R., \& {Livio}, M. 1997, \apj, 491, 312,
  \dodoi{10.1086/304921}

\bibitem[{{Chen} {et~al.}(2022){Chen}, {Sun}, {Li}, {Wang}, {Zhang}, \&
  {Sun}}]{Chen:2022:EP}
{Chen}, Y., {Sun}, X., {Li}, Z., {et~al.} 2022, \ao, 61, 8813,
  \dodoi{10.1364/AO.469433}

\bibitem[{{Coppejans} {et~al.}(2016){Coppejans}, {K{\"o}rding}, {Knigge},
  {Pretorius}, {Woudt}, {Groot}, {Van Eck}, \&
  {Drake}}]{Coppejans:2016:CVstats}
{Coppejans}, D.~L., {K{\"o}rding}, E.~G., {Knigge}, C., {et~al.} 2016, \mnras,
  456, 4441, \dodoi{10.1093/mnras/stv2921}

\bibitem[{{Corral-Santana} {et~al.}(2016{\natexlab{a}}){Corral-Santana},
  {Casares}, {Mu{\~n}oz-Darias}, {Bauer}, {Mart{\'\i}nez-Pais}, \&
  {Russell}}]{Corral-Santana:2016:BlackCAT}
{Corral-Santana}, J.~M., {Casares}, J., {Mu{\~n}oz-Darias}, T., {et~al.}
  2016{\natexlab{a}}, \aap, 587, A61, \dodoi{10.1051/0004-6361/201527130}

\bibitem[{{Corral-Santana} {et~al.}(2016{\natexlab{b}}){Corral-Santana},
  {Casares}, {Mu{\~n}oz-Darias}, {Bauer}, {Mart{\'\i}nez-Pais}, \&
  {Russell}}]{Corral-Santana:15:BlackCAT}
---. 2016{\natexlab{b}}, \aap, 587, A61, \dodoi{10.1051/0004-6361/201527130}

\bibitem[{{Dahiwale} \& {Fremling}(2020)}]{2020qbw:2020:classification}
{Dahiwale}, A., \& {Fremling}, C. 2020, Transient Name Server Classification
  Report, 2020-2302, 1

\bibitem[{{de Martino} {et~al.}(2022){de Martino}, {D'Avanzo}, {Ambrosino},
  {Miraval Zanon}, {Papitto}, {Campana}, {Baglio}, \&
  {Sanna}}]{deMartino:2022:optMAXIJ1826}
{de Martino}, D., {D'Avanzo}, P., {Ambrosino}, F., {et~al.} 2022, The
  Astronomer's Telegram, 15479, 1

\bibitem[{{Dekany} {et~al.}(2020){Dekany}, {Smith}, {Riddle}, {Feeney},
  {Porter}, {Hale}, {Zolkower}, {Belicki}, {Kaye}, {Henning}, {Walters},
  {Cromer}, {Delacroix}, {Rodriguez}, {Reiley}, {Mao}, {Hover}, {Murphy},
  {Burruss}, {Baker}, {Kowalski}, {Reif}, {Mueller}, {Bellm}, {Graham}, \&
  {Kulkarni}}]{Dekany:2020:ObservingSystem}
{Dekany}, R., {Smith}, R.~M., {Riddle}, R., {et~al.} 2020, \pasp, 132, 038001,
  \dodoi{10.1088/1538-3873/ab4ca2}

\bibitem[{{Dubus} {et~al.}(2001){Dubus}, {Hameury}, \&
  {Lasota}}]{Dubus:2001:DIMinXRBs}
{Dubus}, G., {Hameury}, J.~M., \& {Lasota}, J.~P. 2001, \aap, 373, 251,
  \dodoi{10.1051/0004-6361:20010632}

\bibitem[{{Duev} {et~al.}(2019){Duev}, {Mahabal}, {Masci}, {Graham},
  {Rusholme}, {Walters}, {Karmarkar}, {Frederick}, {Kasliwal}, {Rebbapragada},
  \& {Ward}}]{Duev:2019:drb}
{Duev}, D.~A., {Mahabal}, A., {Masci}, F.~J., {et~al.} 2019, \mnras, 489, 3582,
  \dodoi{10.1093/mnras/stz2357}

\bibitem[{{Evans} {et~al.}(2010){Evans}, {Primini}, {Glotfelty}, {Anderson},
  {Bonaventura}, {Chen}, {Davis}, {Doe}, {Evans}, {Fabbiano}, {Galle}, {Gibbs},
  {Grier}, {Hain}, {Hall}, {Harbo}, {(Helen He}, {Houck}, {Karovska},
  {Kashyap}, {Lauer}, {McCollough}, {McDowell}, {Miller}, {Mitschang},
  {Morgan}, {Mossman}, {Nichols}, {Nowak}, {Plummer}, {Refsdal}, {Rots},
  {Siemiginowska}, {Sundheim}, {Tibbetts}, {Van Stone}, {Winkelman}, \&
  {Zografou}}]{Evans:10:CSC}
{Evans}, I.~N., {Primini}, F.~A., {Glotfelty}, K.~J., {et~al.} 2010, {\apjs},
  189, 37

\bibitem[{{Evans} {et~al.}(2020){Evans}, {Page}, {Osborne}, {Beardmore},
  {Willingale}, {Burrows}, {Kennea}, {Perri}, {Capalbi}, {Tagliaferri}, \&
  {Cenko}}]{Evans:20:2SXPS}
{Evans}, P.~A., {Page}, K.~L., {Osborne}, J.~P., {et~al.} 2020, \apjs, 247, 54

\bibitem[{{F{\"o}rster} {et~al.}(2021){F{\"o}rster}, {Cabrera-Vives},
  {Castillo-Navarrete}, {Est{\'e}vez}, {S{\'a}nchez-S{\'a}ez}, {Arredondo},
  {Bauer}, {Carrasco-Davis}, {Catelan}, {Elorrieta}, {Eyheramendy}, {Huijse},
  {Pignata}, {Reyes}, {Reyes}, {Rodr{\'\i}guez-Mancini}, {Ruz-Mieres},
  {Valenzuela}, {{\'A}lvarez-Maldonado}, {Astorga}, {Borissova}, {Clocchiatti},
  {De Cicco}, {Donoso-Oliva}, {Hern{\'a}ndez-Garc{\'\i}a}, {Graham},
  {Jord{\'a}n}, {Kurtev}, {Mahabal}, {Maureira}, {Mu{\~n}oz-Arancibia},
  {Molina-Ferreiro}, {Moya}, {Palma}, {P{\'e}rez-Carrasco}, {Protopapas},
  {Romero}, {Sabatini-Gacitua}, {S{\'a}nchez}, {San Mart{\'\i}n},
  {Sep{\'u}lveda-Cobo}, {Vera}, \& {Vergara}}]{Forster:21:ALeRCE}
{F{\"o}rster}, F., {Cabrera-Vives}, G., {Castillo-Navarrete}, E., {et~al.}
  2021, \aj, 161, 242

\bibitem[{{Gandhi} {et~al.}(2020){Gandhi}, {Rao}, {Charles}, {Belczynski},
  {Maccarone}, {Arur}, \& {Corral-Santana}}]{Gandhi:2020:LMXBspatial}
{Gandhi}, P., {Rao}, A., {Charles}, P.~A., {et~al.} 2020, \mnras, 496, L22,
  \dodoi{10.1093/mnrasl/slaa081}

\bibitem[{{Graham} {et~al.}(2019){Graham}, {Kulkarni}, {Bellm}, {Adams},
  {Barbarino}, {Blagorodnova}, {Bodewits}, {Bolin}, {Brady}, {Cenko}, {Chang},
  {Coughlin}, {De}, {Eadie}, {Farnham}, {Feindt}, {Franckowiak}, {Fremling},
  {Gezari}, {Ghosh}, {Goldstein}, {Golkhou}, {Goobar}, {Ho}, {Huppenkothen},
  {Ivezi{\'c}}, {Jones}, {Juric}, {Kaplan}, {Kasliwal}, {Kelley}, {Kupfer},
  {Lee}, {Lin}, {Lunnan}, {Mahabal}, {Miller}, {Ngeow}, {Nugent}, {Ofek},
  {Prince}, {Rauch}, {van Roestel}, {Schulze}, {Singer}, {Sollerman}, {Taddia},
  {Yan}, {Ye}, {Yu}, {Barlow}, {Bauer}, {Beck}, {Belicki}, {Biswas}, {Brinnel},
  {Brooke}, {Bue}, {Bulla}, {Burruss}, {Connolly}, {Cromer}, {Cunningham},
  {Dekany}, {Delacroix}, {Desai}, {Duev}, {Feeney}, {Flynn}, {Frederick},
  {Gal-Yam}, {Giomi}, {Groom}, {Hacopians}, {Hale}, {Helou}, {Henning},
  {Hover}, {Hillenbrand}, {Howell}, {Hung}, {Imel}, {Ip}, {Jackson}, {Kaspi},
  {Kaye}, {Kowalski}, {Kramer}, {Kuhn}, {Landry}, {Laher}, {Mao}, {Masci},
  {Monkewitz}, {Murphy}, {Nordin}, {Patterson}, {Penprase}, {Porter},
  {Rebbapragada}, {Reiley}, {Riddle}, {Rigault}, {Rodriguez}, {Rusholme}, {van
  Santen}, {Shupe}, {Smith}, {Soumagnac}, {Stein}, {Surace}, {Szkody}, {Terek},
  {Van Sistine}, {van Velzen}, {Vestrand}, {Walters}, {Ward}, {Zhang}, \&
  {Zolkower}}]{Graham:2019:ZTFscience}
{Graham}, M.~J., {Kulkarni}, S.~R., {Bellm}, E.~C., {et~al.} 2019, \pasp, 131,
  078001, \dodoi{10.1088/1538-3873/ab006c}

\bibitem[{{Green} {et~al.}(2019){Green}, {Schlafly}, {Zucker}, {Speagle}, \&
  {Finkbeiner}}]{Green:19:3DDustMap}
{Green}, G.~M., {Schlafly}, E., {Zucker}, C., {Speagle}, J.~S., \&
  {Finkbeiner}, D. 2019, \apj, 887, 93, \dodoi{10.3847/1538-4357/ab5362}

\bibitem[{{Grimm} {et~al.}(2002){Grimm}, {Gilfanov}, \&
  {Sunyaev}}]{Grimm:02:XRBDistribution}
{Grimm}, H.~J., {Gilfanov}, M., \& {Sunyaev}, R. 2002, \aap, 391, 923,
  \dodoi{10.1051/0004-6361:20020826}

\bibitem[{Hameury(2020)}]{Ham:2020:DIMReview}
Hameury, J. 2020, Advances in Space Research, 66, 1004,
  \dodoi{https://doi.org/10.1016/j.asr.2019.10.022}

\bibitem[{{Heinke} {et~al.}(2015){Heinke}, {Bahramian}, {Degenaar}, \&
  {Wijnands}}]{Heinke:2015:VFXTs}
{Heinke}, C.~O., {Bahramian}, A., {Degenaar}, N., \& {Wijnands}, R. 2015,
  \mnras, 447, 3034, \dodoi{10.1093/mnras/stu2652}

\bibitem[{{Ivezi{\'c}} {et~al.}(2019){Ivezi{\'c}}, {Kahn}, {Tyson}, {Abel},
  {Acosta}, {Allsman}, {Alonso}, {AlSayyad}, {Anderson}, {Andrew}, {Angel},
  {Angeli}, {Ansari}, {Antilogus}, {Araujo}, {Armstrong}, {Arndt}, {Astier},
  {Aubourg}, {Auza}, {Axelrod}, {Bard}, {Barr}, {Barrau}, {Bartlett}, {Bauer},
  {Bauman}, {Baumont}, {Bechtol}, {Bechtol}, {Becker}, {Becla}, {Beldica},
  {Bellavia}, {Bianco}, {Biswas}, {Blanc}, {Blazek}, {Blandford}, {Bloom},
  {Bogart}, {Bond}, {Booth}, {Borgland}, {Borne}, {Bosch}, {Boutigny},
  {Brackett}, {Bradshaw}, {Brandt}, {Brown}, {Bullock}, {Burchat}, {Burke},
  {Cagnoli}, {Calabrese}, {Callahan}, {Callen}, {Carlin}, {Carlson},
  {Chandrasekharan}, {Charles-Emerson}, {Chesley}, {Cheu}, {Chiang}, {Chiang},
  {Chirino}, {Chow}, {Ciardi}, {Claver}, {Cohen-Tanugi}, {Cockrum}, {Coles},
  {Connolly}, {Cook}, {Cooray}, {Covey}, {Cribbs}, {Cui}, {Cutri}, {Daly},
  {Daniel}, {Daruich}, {Daubard}, {Daues}, {Dawson}, {Delgado}, {Dellapenna},
  {de Peyster}, {de Val-Borro}, {Digel}, {Doherty}, {Dubois},
  {Dubois-Felsmann}, {Durech}, {Economou}, {Eifler}, {Eracleous}, {Emmons},
  {Fausti Neto}, {Ferguson}, {Figueroa}, {Fisher-Levine}, {Focke}, {Foss},
  {Frank}, {Freemon}, {Gangler}, {Gawiser}, {Geary}, {Gee}, {Geha}, {Gessner},
  {Gibson}, {Gilmore}, {Glanzman}, {Glick}, {Goldina}, {Goldstein}, {Goodenow},
  {Graham}, {Gressler}, {Gris}, {Guy}, {Guyonnet}, {Haller}, {Harris},
  {Hascall}, {Haupt}, {Hernandez}, {Herrmann}, {Hileman}, {Hoblitt}, {Hodgson},
  {Hogan}, {Howard}, {Huang}, {Huffer}, {Ingraham}, {Innes}, {Jacoby}, {Jain},
  {Jammes}, {Jee}, {Jenness}, {Jernigan}, {Jevremovi{\'c}}, {Johns}, {Johnson},
  {Johnson}, {Jones}, {Juramy-Gilles}, {Juri{\'c}}, {Kalirai}, {Kallivayalil},
  {Kalmbach}, {Kantor}, {Karst}, {Kasliwal}, {Kelly}, {Kessler}, {Kinnison},
  {Kirkby}, {Knox}, {Kotov}, {Krabbendam}, {Krughoff}, {Kub{\'a}nek},
  {Kuczewski}, {Kulkarni}, {Ku}, {Kurita}, {Lage}, {Lambert}, {Lange},
  {Langton}, {Le Guillou}, {Levine}, {Liang}, {Lim}, {Lintott}, {Long},
  {Lopez}, {Lotz}, {Lupton}, {Lust}, {MacArthur}, {Mahabal}, {Mandelbaum},
  {Markiewicz}, {Marsh}, {Marshall}, {Marshall}, {May}, {McKercher}, {McQueen},
  {Meyers}, {Migliore}, {Miller}, {Mills}, {Miraval}, {Moeyens}, {Moolekamp},
  {Monet}, {Moniez}, {Monkewitz}, {Montgomery}, {Morrison}, {Mueller},
  {Muller}, {Mu{\~n}oz Arancibia}, {Neill}, {Newbry}, {Nief}, {Nomerotski},
  {Nordby}, {O'Connor}, {Oliver}, {Olivier}, {Olsen}, {O'Mullane}, {Ortiz},
  {Osier}, {Owen}, {Pain}, {Palecek}, {Parejko}, {Parsons}, {Pease},
  {Peterson}, {Peterson}, {Petravick}, {Libby Petrick}, {Petry},
  {Pierfederici}, {Pietrowicz}, {Pike}, {Pinto}, {Plante}, {Plate}, {Plutchak},
  {Price}, {Prouza}, {Radeka}, {Rajagopal}, {Rasmussen}, {Regnault}, {Reil},
  {Reiss}, {Reuter}, {Ridgway}, {Riot}, {Ritz}, {Robinson}, {Roby}, {Roodman},
  {Rosing}, {Roucelle}, {Rumore}, {Russo}, {Saha}, {Sassolas}, {Schalk},
  {Schellart}, {Schindler}, {Schmidt}, {Schneider}, {Schneider}, {Schoening},
  {Schumacher}, {Schwamb}, {Sebag}, {Selvy}, {Sembroski}, {Seppala}, {Serio},
  {Serrano}, {Shaw}, {Shipsey}, {Sick}, {Silvestri}, {Slater}, {Smith},
  {Smith}, {Sobhani}, {Soldahl}, {Storrie-Lombardi}, {Stover}, {Strauss},
  {Street}, {Stubbs}, {Sullivan}, {Sweeney}, {Swinbank}, {Szalay}, {Takacs},
  {Tether}, {Thaler}, {Thayer}, {Thomas}, {Thornton}, {Thukral}, {Tice},
  {Trilling}, {Turri}, {Van Berg}, {Vanden Berk}, {Vetter}, {Virieux},
  {Vucina}, {Wahl}, {Walkowicz}, {Walsh}, {Walter}, {Wang}, {Wang}, {Warner},
  {Wiecha}, {Willman}, {Winters}, {Wittman}, {Wolff}, {Wood-Vasey}, {Wu},
  {Xin}, {Yoachim}, \& {Zhan}}]{Ivezic:2019:LSST}
{Ivezi{\'c}}, {\v{Z}}., {Kahn}, S.~M., {Tyson}, J.~A., {et~al.} 2019, \apj,
  873, 111, \dodoi{10.3847/1538-4357/ab042c}

\bibitem[{{Johnson} {et~al.}(2019){Johnson}, {Gandhi}, {Chapman}, {Moreau},
  {Charles}, {Clarkson}, \& {Hill}}]{Johnson:19:LSSTLMXB}
{Johnson}, M. A.~C., {Gandhi}, P., {Chapman}, A.~P., {et~al.} 2019, \mnras,
  484, 19, \dodoi{10.1093/mnras/sty3466}

\bibitem[{{King}(1998)}]{King:1998:Irr}
{King}, A.~R. 1998, \mnras, 296, L45, \dodoi{10.1046/j.1365-8711.1998.01652.x}

\bibitem[{{King} \& {Ritter}(1998)}]{King:1998:SoftX-rayLCs}
{King}, A.~R., \& {Ritter}, H. 1998, \mnras, 293, L42,
  \dodoi{10.1046/j.1365-8711.1998.01295.x}

\bibitem[{Krimm {et~al.}(2013)Krimm, Holland, Corbet, Pearlman, Romano, Kennea,
  Bloom, Barthelmy, Baumgartner, Cummings, Gehrels, Lien, Markwardt, Palmer,
  Sakamoto, Stamatikos, \& Ukwatta}]{Krimm:2013:Swift_BAT}
Krimm, H.~A., Holland, S.~T., Corbet, R. H.~D., {et~al.} 2013, The
  Astrophysical Journal Supplement Series, 209, 14,
  \dodoi{10.1088/0067-0049/209/1/14}

\bibitem[{{Krimm} {et~al.}(2018){Krimm}, {Barthelmy}, {Cummings}, {Lien},
  {Markwardt}, {Palmer}, {Sakamoto}, {Stamatikos}, \&
  {Ukwatta}}]{Krimm:2018:atel_swiftJ1858}
{Krimm}, H.~A., {Barthelmy}, S.~D., {Cummings}, J.~R., {et~al.} 2018, The
  Astronomer's Telegram, 12151, 1

\bibitem[{{Lallement} {et~al.}(2019){Lallement}, {Babusiaux}, {Vergely},
  {Katz}, {Arenou}, {Valette}, {Hottier}, \& {Capitanio}}]{Lallement:19:3DDust}
{Lallement}, R., {Babusiaux}, C., {Vergely}, J.~L., {et~al.} 2019, \aap, 625,
  A135, \dodoi{10.1051/0004-6361/201834695}

\bibitem[{{Martini} {et~al.}(2011){Martini}, {Stoll}, {Derwent}, {Zhelem},
  {Atwood}, {Gonzalez}, {Mason}, {O'Brien}, {Pappalardo}, {Pogge}, {Ward}, \&
  {Wong}}]{OSMOS:2011}
{Martini}, P., {Stoll}, R., {Derwent}, M.~A., {et~al.} 2011, \pasp, 123, 187,
  \dodoi{10.1086/658357}

\bibitem[{{Masci} {et~al.}(2019){Masci}, {Laher}, {Rusholme}, {Shupe}, {Groom},
  {Surace}, {Jackson}, {Monkewitz}, {Beck}, {Flynn}, {Terek}, {Landry},
  {Hacopians}, {Desai}, {Howell}, {Brooke}, {Imel}, {Wachter}, {Ye}, {Lin},
  {Cenko}, {Cunningham}, {Rebbapragada}, {Bue}, {Miller}, {Mahabal}, {Bellm},
  {Patterson}, {Juri{\'c}}, {Golkhou}, {Ofek}, {Walters}, {Graham}, {Kasliwal},
  {Dekany}, {Kupfer}, {Burdge}, {Cannella}, {Barlow}, {Van Sistine}, {Giomi},
  {Fremling}, {Blagorodnova}, {Levitan}, {Riddle}, {Smith}, {Helou}, {Prince},
  \& {Kulkarni}}]{Masci:2019:ZTFpipeline}
{Masci}, F.~J., {Laher}, R.~R., {Rusholme}, B., {et~al.} 2019, \pasp, 131,
  018003, \dodoi{10.1088/1538-3873/aae8ac}

\bibitem[{{Matheson} {et~al.}(2021){Matheson}, {Stubens}, {Wolf}, {Lee},
  {Narayan}, {Saha}, {Scott}, {Soraisam}, {Bolton}, {Hauger}, {Silva},
  {Kececioglu}, {Scheidegger}, {Snodgrass}, {Aleo}, {Evans-Jacquez}, {Singh},
  {Wang}, {Yang}, \& {Zhao}}]{Matheson:21:ANTARES}
{Matheson}, T., {Stubens}, C., {Wolf}, N., {et~al.} 2021, \aj, 161, 107

\bibitem[{{Matsuoka} {et~al.}(2009){Matsuoka}, {Kawasaki}, {Ueno}, {Tomida},
  {Kohama}, {Suzuki}, {Adachi}, {Ishikawa}, {Mihara}, {Sugizaki}, {Isobe},
  {Nakagawa}, {Tsunemi}, {Miyata}, {Kawai}, {Kataoka}, {Morii}, {Yoshida},
  {Negoro}, {Nakajima}, {Ueda}, {Chujo}, {Yamaoka}, {Yamazaki}, {Nakahira},
  {You}, {Ishiwata}, {Miyoshi}, {Eguchi}, {Hiroi}, {Katayama}, \&
  {Ebisawa}}]{Matsuoka:2009:MAXI}
{Matsuoka}, M., {Kawasaki}, K., {Ueno}, S., {et~al.} 2009, \pasj, 61, 999,
  \dodoi{10.1093/pasj/61.5.999}

\bibitem[{{Mereminskiy} {et~al.}(2020){Mereminskiy}, {Medvedev}, {Semena},
  {Pavlinsky}, {Molkov}, {Lutovinov}, {Burenin}, {Sazonov}, {Sunyaev}, \&
  {Gilfanov}}]{Atel:2020:SRGdetAT2019wey}
{Mereminskiy}, I., {Medvedev}, P., {Semena}, A., {et~al.} 2020, The
  Astronomer's Telegram, 13571, 1

\bibitem[{{M{\"o}ller} {et~al.}(2021){M{\"o}ller}, {Peloton}, {Ishida},
  {Arnault}, {Bachelet}, {Blaineau}, {Boutigny}, {Chauhan}, {Gangler},
  {Hernandez}, {Hrivnac}, {Leoni}, {Leroy}, {Moniez}, {Pateyron}, {Ramparison},
  {Turpin}, {Ansari}, {Allam}, {Bajat}, {Biswas}, {Boucaud}, {Bregeon},
  {Campagne}, {Cohen-Tanugi}, {Coleiro}, {Dornic}, {Fouchez}, {Godet}, {Gris},
  {Karpov}, {Nebot Gomez-Moran}, {Neveu}, {Plaszczynski}, {Savchenko}, \&
  {Webb}}]{Moller:21:FINK}
{M{\"o}ller}, A., {Peloton}, J., {Ishida}, E. E.~O., {et~al.} 2021, \mnras,
  501, 3272

\bibitem[{{Negoro} {et~al.}(2022){Negoro}, {Iwakiri}, {Kawakubo}, {Nakajima},
  {Kobayashi}, {Tanaka}, {Soejima}, {Mihara}, {Kawamuro}, {Yamada}, {Tamagawa},
  {Matsuoka}, {Sakamoto}, {Serino}, {Sugita}, {Hiramatsu}, {Yoshida}, {Tsuboi},
  {Kohara}, {Shidatsu}, {Iwasaki}, {Kawai}, {Niwano}, {Hosokawa}, {Imai},
  {Ito}, {Takamatsu}, {Nakahira}, {Ueno}, {Tomida}, {Ishikawa}, {Tominaga},
  {Nagatsuka}, {Kurihara}, {Ueda}, {Ogawa}, {Setoguchi}, {Yoshitake}, {Inaba},
  {Tsunemi}, {Yamauchi}, {Sato}, {Hatsuda}, {Fukuoka}, {hagiwara}, {Umeki},
  {Yamaoka}, \& {Sugizaki}}]{2022ATel15440....1N}
{Negoro}, H., {Iwakiri}, W., {Kawakubo}, Y., {et~al.} 2022, The Astronomer's
  Telegram, 15440, 1

\bibitem[{{Neilsen} \& {Degenaar}(2023)}]{tmp_Neilsen:23:XRBWindReview}
{Neilsen}, J., \& {Degenaar}, N. 2023, arXiv e-prints, arXiv:2304.05412

\bibitem[{{Niwano} {et~al.}(2023){Niwano}, {Murata}, {Ito}, {Yatsu}, \&
  {Kawai}}]{Niwano:2023:aqlx1}
{Niwano}, M., {Murata}, K.~L., {Ito}, N., {Yatsu}, Y., \& {Kawai}, N. 2023,
  \mnras, 525, 4358, \dodoi{10.1093/mnras/stad2561}

\bibitem[{{Nordin} {et~al.}(2019){Nordin}, {Brinnel}, {van Santen}, {Bulla},
  {Feindt}, {Franckowiak}, {Fremling}, {Gal-Yam}, {Giomi}, {Kowalski},
  {Mahabal}, {Miranda}, {Rauch}, {Rigault}, {Schulze}, {Sollerman}, {Stein},
  {Yaron}, {van Velzen}, \& {Ward}}]{Nordin:19:AMPEL}
{Nordin}, J., {Brinnel}, V., {van Santen}, J., {et~al.} 2019, arXiv e-prints

\bibitem[{{Patterson} {et~al.}(2019){Patterson}, {Bellm}, {Rusholme}, {Masci},
  {Juric}, {Krughoff}, {Golkhou}, {Graham}, {Kulkarni}, {Helou}, \& {Zwicky
  Transient Facility Collaboration}}]{Pat19}
{Patterson}, M.~T., {Bellm}, E.~C., {Rusholme}, B., {et~al.} 2019, \pasp, 131,
  018001, \dodoi{10.1088/1538-3873/aae904}

\bibitem[{{Pirbhoy} {et~al.}(2020){Pirbhoy}, {Baglio}, {Russell}, {Bramich},
  {Saikia}, {Yazeedi}, \& {Lewis}}]{XBnews:2020:detMAXIJ1348}
{Pirbhoy}, S.~F., {Baglio}, M.~C., {Russell}, D.~M., {et~al.} 2020, The
  Astronomer's Telegram, 13451, 1

\bibitem[{{Priedhorsky} \& {Verbunt}(1988)}]{Priedhorsky:1988:TidalXRBs}
{Priedhorsky}, W.~C., \& {Verbunt}, F. 1988, \apj, 333, 895,
  \dodoi{10.1086/166798}

\bibitem[{{Rau} {et~al.}(2012){Rau}, {Nardini}, \& {Greiner}}]{atel:2012:GROND}
{Rau}, A., {Nardini}, M., \& {Greiner}, J. 2012, The Astronomer's Telegram,
  4054, 1

\bibitem[{{Reig} \& {Fabregat}(2015)}]{Reig:2015:OptHMXBs}
{Reig}, P., \& {Fabregat}, J. 2015, \aap, 574, A33,
  \dodoi{10.1051/0004-6361/201425008}

\bibitem[{{Ritter} \& {Kolb}(2003)}]{Ritter:2003:XRBcat}
{Ritter}, H., \& {Kolb}, U. 2003, \aap, 404, 301,
  \dodoi{10.1051/0004-6361:20030330}

\bibitem[{{Russell} {et~al.}(2019){Russell}, {Bramich}, {Lewis}, {AlMannaei},
  {Al Qaissieh}, {Al Qasim}, {Al Yazeedi}, {Baglio}, {Bernardini}, {Elgalad},
  {Gabuya}, {Lasota}, {Palado}, {Roche}, {Shivkumar}, {Udrescu}, \&
  {Zhang}}]{Rus19}
{Russell}, D.~M., {Bramich}, D.~M., {Lewis}, F., {et~al.} 2019, Astronomische
  Nachrichten, 340, 278, \dodoi{10.1002/asna.201913610}

\bibitem[{Saikia {et~al.}(2023)Saikia, Russell, Pirbhoy, Baglio, Bramich,
  Alabarta, Lewis, \& Charles}]{Saikia:2023:OptMonBXBH}
Saikia, P., Russell, D.~M., Pirbhoy, S.~F., {et~al.} 2023, The Astrophysical
  Journal, 949, 104, \dodoi{10.3847/1538-4357/acc8cc}

\bibitem[{{Sanna} {et~al.}(2022){Sanna}, {Bult}, {Ng}, {Ray}, {Jaisawal},
  {Burderi}, {Di Salvo}, {Riggio}, {Altamirano}, {Strohmayer}, {Manca},
  {Gendreau}, {Chakrabarty}, {Iwakiri}, \& {Iaria}}]{Sanna:2022:AMXP}
{Sanna}, A., {Bult}, P., {Ng}, M., {et~al.} 2022, \mnras, 516, L76,
  \dodoi{10.1093/mnrasl/slac093}

\bibitem[{{Shakura} \& {Sunyaev}(1973)}]{Shakura:1973}
{Shakura}, N.~I., \& {Sunyaev}, R.~A. 1973, \aap, 24, 337

\bibitem[{{Shidatsu} {et~al.}(2019){Shidatsu}, {Negoro}, {Nakajima},
  {Maruyama}, {Aoki}, {Kobayashi}, {Mihara}, {Tamagawa}, {Matsuoka},
  {Sakamoto}, {Serino}, {Sugita}, {Nishida}, {Yoshida}, {Tsuboi}, {Iwakiri},
  {Sasaki}, {Kawai}, {Sato}, {Kawai}, {Sugizaki}, {Oeda}, {Shiraishi},
  {Nakahira}, {Sugawara}, {Ueno}, {Tomida}, {Ishikawa}, {Isobe}, {Shimomukai},
  {Tominaga}, {Ueda}, {Tanimoto}, {Yamada}, {Ogawa}, {Setoguchi}, {Yoshitake},
  {Tsunemi}, {Yoneyama}, {Asakura}, {Ide}, {Yamauchi}, {Iwahori}, {Kurihara},
  {Kurogi}, {Miike}, {Kawamuro}, {Yamaoka}, \& {Kawakubo}}]{atel:MAXIJ1807+132}
{Shidatsu}, M., {Negoro}, H., {Nakajima}, M., {et~al.} 2019, The Astronomer's
  Telegram, 13097, 1

\bibitem[{{Smith} {et~al.}(2019){Smith}, {Williams}, {Young}, {Ibsen},
  {Smartt}, {Lawrence}, {Morris}, {Voutsinas}, \& {Nicholl}}]{Smith:19:Lasair}
{Smith}, K.~W., {Williams}, R.~D., {Young}, D.~R., {et~al.} 2019, Research
  Notes of the American Astronomical Society, 3, 26

\bibitem[{Tetarenko {et~al.}(2018)Tetarenko, Dubus, Lasota, Heinke, \&
  Sivakoff}]{Tetarenko:2018:XrayIrr}
Tetarenko, B., Dubus, G., Lasota, J.-P., Heinke, C., \& Sivakoff, G. 2018,
  Monthly Notices of the Royal Astronomical Society, 480, 2,
  \dodoi{10.1093/mnras/sty1798}

\bibitem[{Tetarenko {et~al.}(2016)Tetarenko, Sivakoff, Heinke, \&
  Gladstone}]{Tet16}
Tetarenko, B.~E., Sivakoff, G.~R., Heinke, C.~O., \& Gladstone, J.~C. 2016, The
  Astrophysical Journal Supplement Series, 222, 15,
  \dodoi{10.3847/0067-0049/222/2/15}

\bibitem[{{The Rubin Observatory Survey Cadence Optimization
  Committee}(2023)}]{PSTN-055}
{The Rubin Observatory Survey Cadence Optimization Committee}. 2023, {Survey
  Cadence Optimization Committee’s Phase 2 Recommendations},
  \url{ls.st/PSTN-055}

\bibitem[{{Tonry} {et~al.}(2019){Tonry}, {Denneau}, {Heinze}, {Weiland},
  {Flewelling}, {Stalder}, {Rest}, {Stubbs}, {Smith}, {Smartt}, {Young},
  {Srivastav}, {McBrien}, {O'Neill}, {Clark}, {Fulton}, {Gillanders}, {Dobson},
  {Chen}, {Wright}, \& {Anderson}}]{TNS:2019:ATLASdet2019wey}
{Tonry}, J., {Denneau}, L., {Heinze}, A., {et~al.} 2019, Transient Name Server
  Discovery Report, 2019-2553, 1

\bibitem[{{Tucker} {et~al.}(2018){Tucker}, {Shappee}, {Holoien}, {Auchettl},
  {Strader}, {Stanek}, {Kochanek}, {Bahramian}, {ASAS-SN}, {Dong}, {Prieto},
  {Shields}, {Thompson}, {Beacom}, {Chomiuk}, {ATLAS}, {Denneau}, {Flewelling},
  {Heinze}, {Smith}, {Stalder}, {Tonry}, {Weiland}, {Rest}, {Huber}, {Rowan},
  \& {Dage}}]{Tucker:18:ASASSN-18ey}
{Tucker}, M.~A., {Shappee}, B.~J., {Holoien}, T.~W.-S., {et~al.} 2018, \apjl,
  867, L9

\bibitem[{{Wang} {et~al.}(2022){Wang}, {Bellm}, \&
  {Jaodand}}]{2022ATel15455....1W}
{Wang}, Y., {Bellm}, E.~C., \& {Jaodand}, A. 2022, The Astronomer's Telegram,
  15455, 1

\bibitem[{{Warwick} {et~al.}(2012){Warwick}, {Saxton}, \&
  {Read}}]{Warwick:12:XMMSlewSurvey}
{Warwick}, R.~S., {Saxton}, R.~D., \& {Read}, A.~M. 2012, \aap, 548, A99

\bibitem[{{Webb} {et~al.}(2020){Webb}, {Coriat}, {Traulsen}, {Ballet}, {Motch},
  {Carrera}, {Koliopanos}, {Authier}, {de la Calle}, {Ceballos}, {Colomo},
  {Chuard}, {Freyberg}, {Garcia}, {Kolehmainen}, {Lamer}, {Lin}, {Maggi},
  {Michel}, {Page}, {Page}, {Perea-Calderon}, {Pineau}, {Rodriguez}, {Rosen},
  {Santos Lleo}, {Saxton}, {Schwope}, {Tom{\'a}s}, {Watson}, \&
  {Zakardjian}}]{Webb:20:4XMMDR10}
{Webb}, N.~A., {Coriat}, M., {Traulsen}, I., {et~al.} 2020, \aap, 641, A136

\bibitem[{{Yao} {et~al.}(2021){Yao}, {Kulkarni}, {Burdge}, {Caiazzo}, {De},
  {Dong}, {Fremling}, {Kasliwal}, {Kupfer}, {van Roestel}, {Sollerman},
  {Bagdasaryan}, {Bellm}, {Cenko}, {Drake}, {Duev}, {Graham}, {Kaye}, {Masci},
  {Miranda}, {Prince}, {Riddle}, {Rusholme}, \& {Soumagnac}}]{Yao21}
{Yao}, Y., {Kulkarni}, S.~R., {Burdge}, K.~B., {et~al.} 2021, \apj, 920, 120,
  \dodoi{10.3847/1538-4357/ac15f9}

\end{thebibliography}

\appendix 
\section{ X-ray and Optical Lightcurves of LMXB Outbursts }

\begin{figure}
    \centering
    \includegraphics[width=0.45\textwidth]{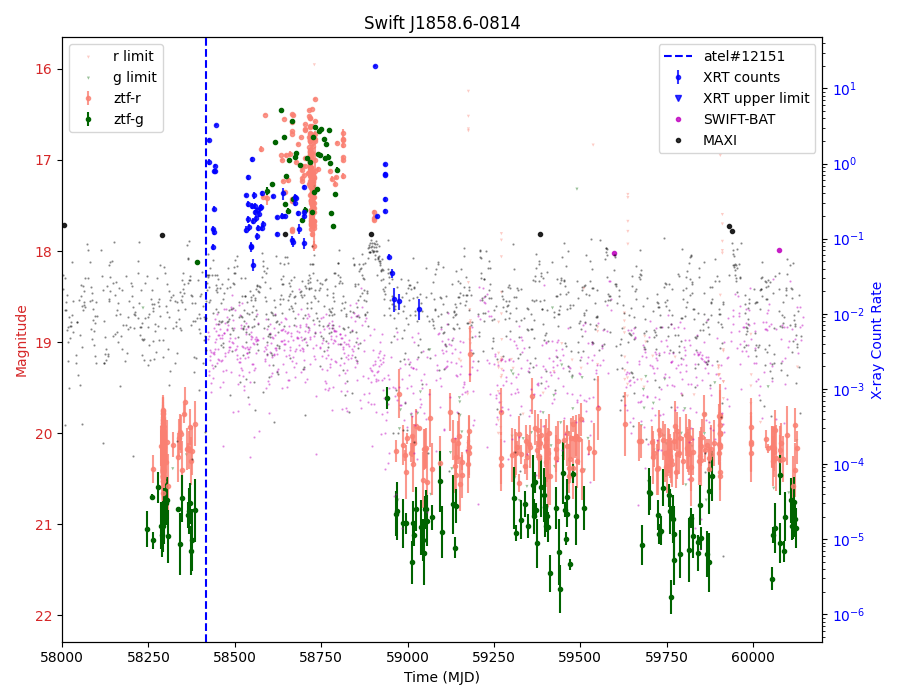}
    \includegraphics[width=0.45\textwidth]{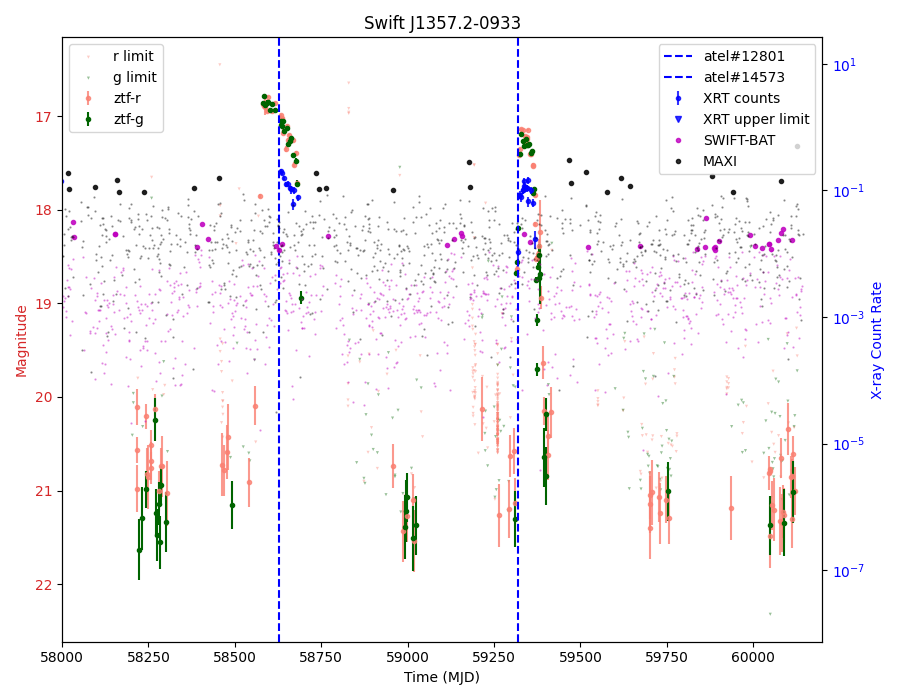}
    \includegraphics[width=0.45\textwidth]{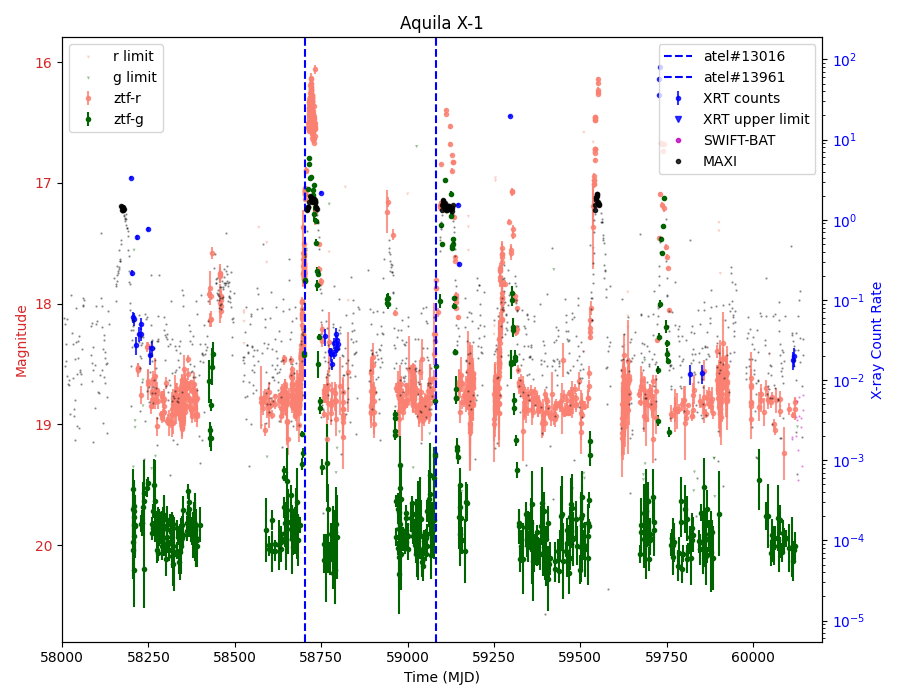}
    \includegraphics[width=0.45\textwidth]{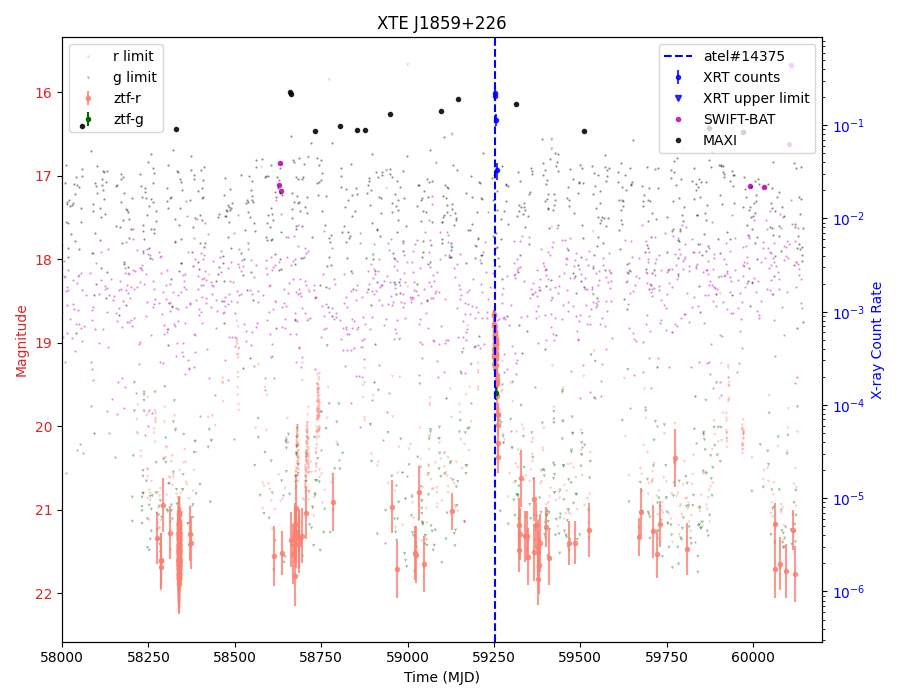}
    \includegraphics[width=0.45\textwidth]{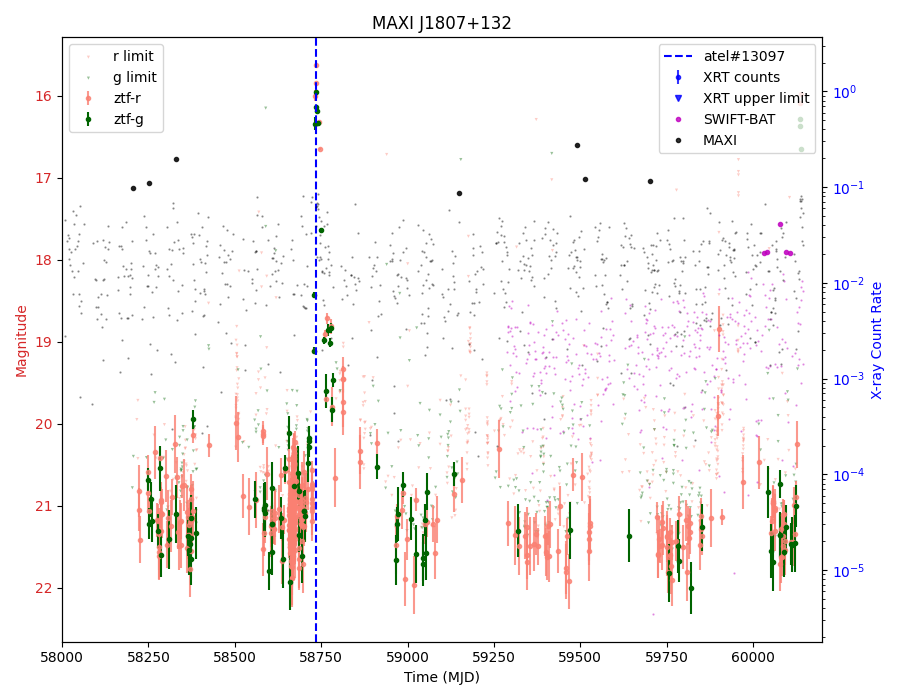}
    \includegraphics[width=0.45\textwidth]{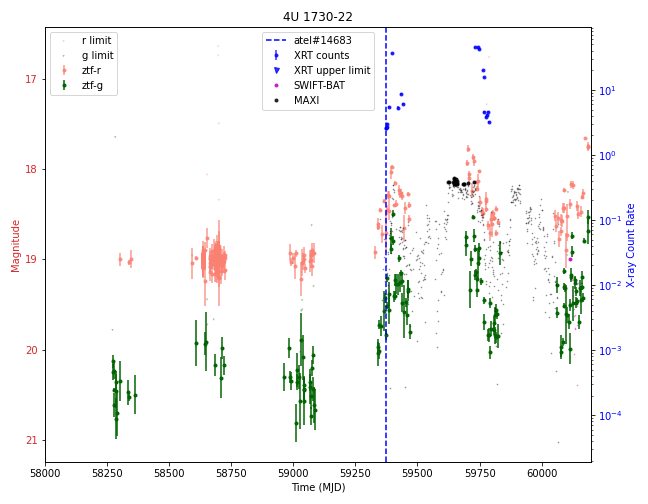}

    \caption{\blue{ZTF} optical and X-ray lightcurves of LMXBs that have \blue{been reported in} outburst since \blue{June 2018}. These 6 sources all have at least one ZTF observation prior to the date and time of the first observation by an all-sky monitor as reported in the atel in the legend. The red and green points are ZTF magnitudes in r and g band. The blue points give counts between 2--10\,keV as detected by Swift XRT. The back points give counts observed by Swift BAT (15--150\,keV) and the purple points are MAXI observations (2--30\,keV). Larger BAT and MAXI points \rev{are detections exceeding 3$\sigma$.  Swift J1858.6-0814 was initially discovered in 2018 \citep{Krimm:2018:atel_swiftJ1858}, meaning this source could have been detected by ZTF as a then unknown LMXB before X-ray ASMs.}}
    \label{fig:optfirst}
\end{figure}

\begin{figure}
    \centering
    \includegraphics[width=0.45\textwidth]{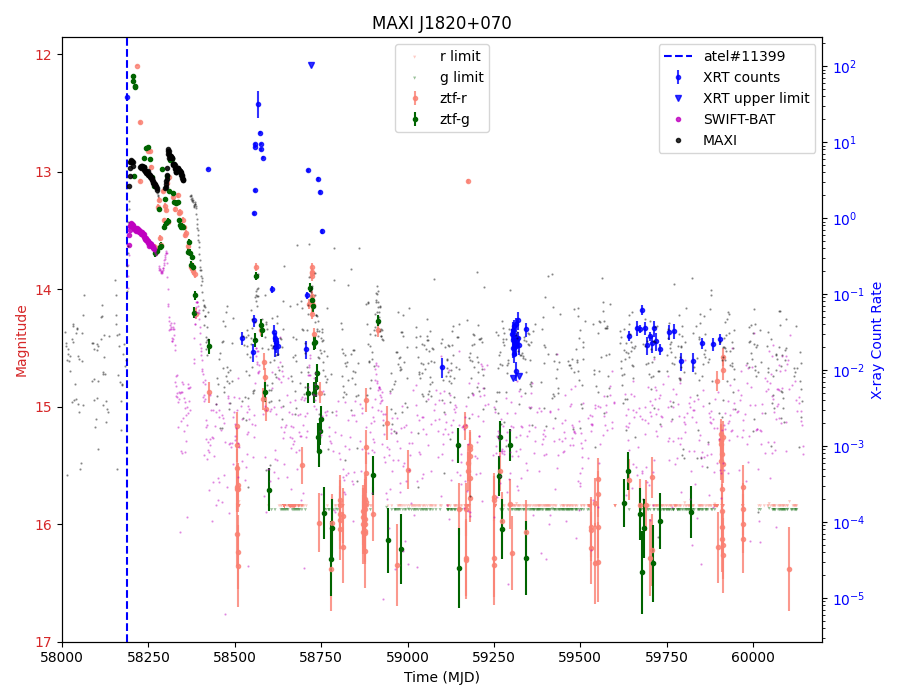}
    \includegraphics[width=0.45\textwidth]{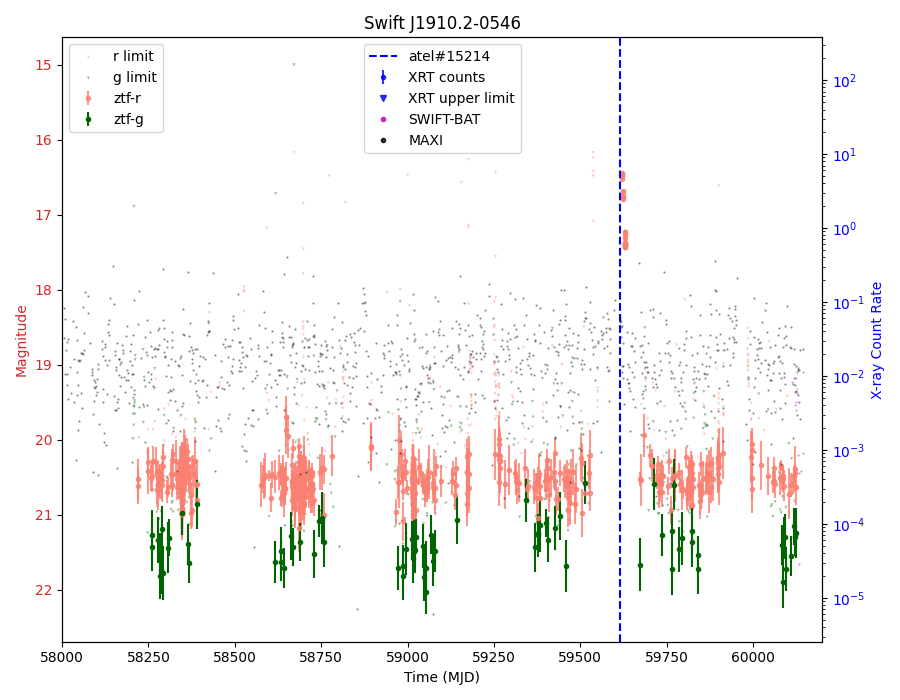}
    \includegraphics[width=0.45\textwidth]{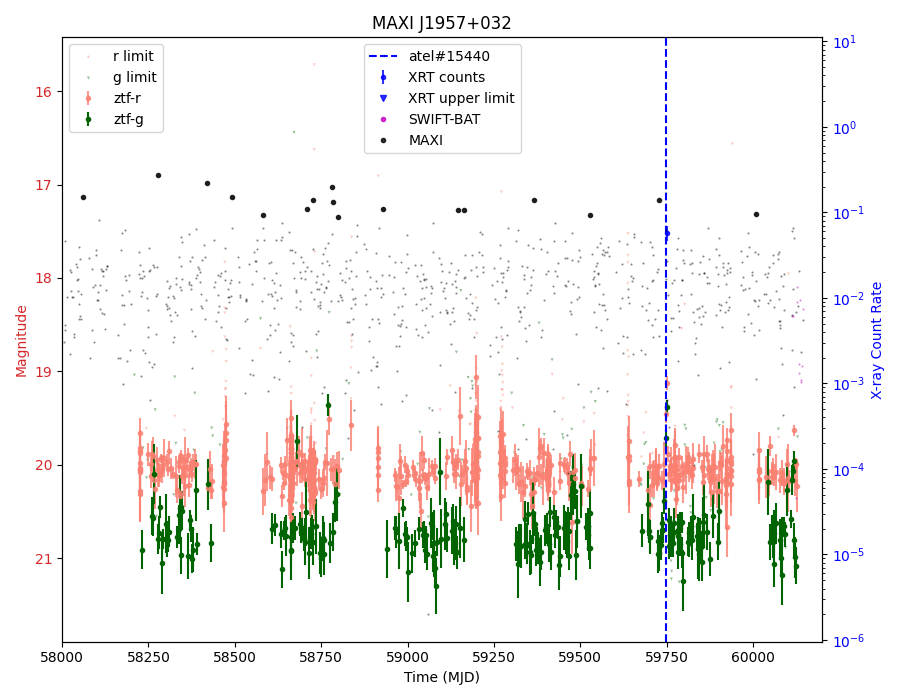}

    \caption{Optical and X-ray lightcurves of LMXBs that have gone into outburst since the beginning of ZTF. These 3 sources have simultaneous X-ray and ZTF data. \rev{The first discovery of MAXI J1820 was in the optical by ASAS-SN \citep{Tucker:18:ASASSN-18ey}; ZTF did not observe MAXI J1820 prior to the X-ray detections.}}
    \label{fig:optxray}
\end{figure}

\begin{figure}
    \centering
    \includegraphics[width=0.45\textwidth]{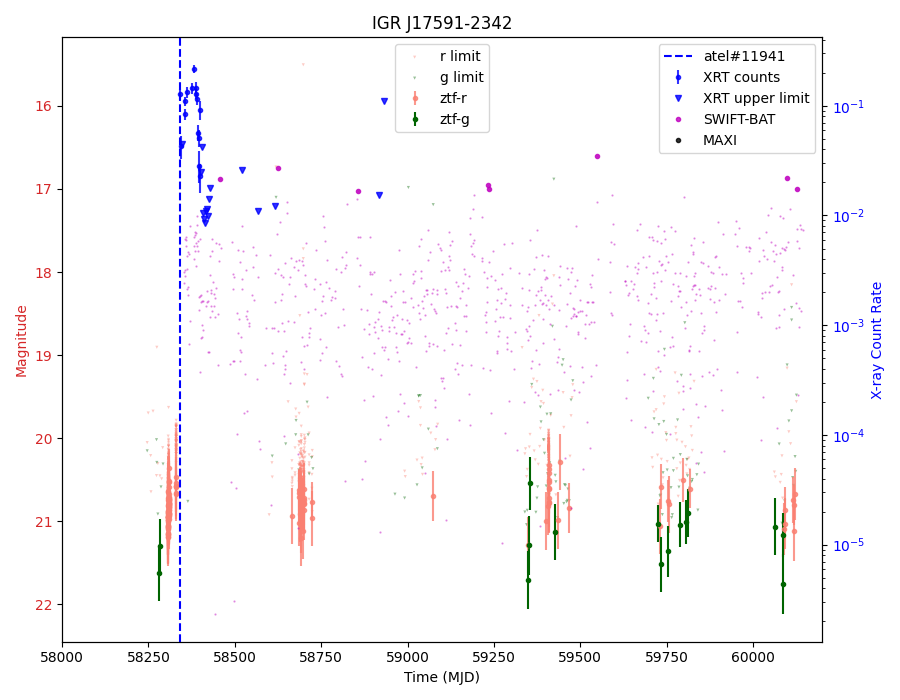}
    \includegraphics[width=0.45\textwidth]{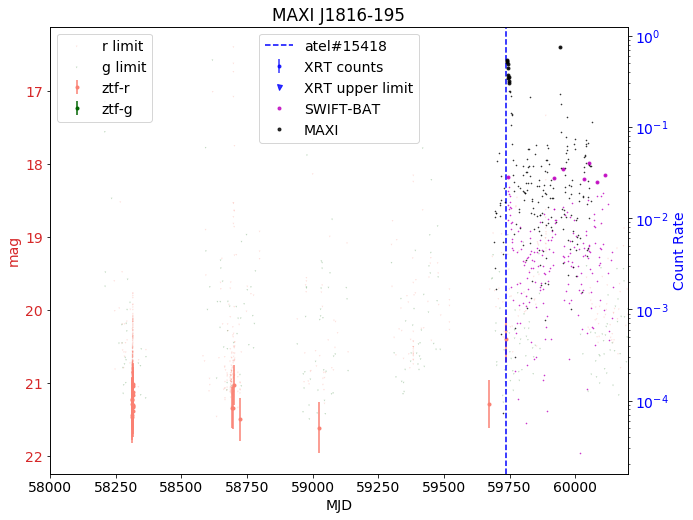}

    \caption{Optical and X-ray lightcurves of an LMXB that have gone into outburst since the beginning of ZTF. These sources display outbursts in X-ray but no available or inconclusive optical brightening in ZTF data. MAXI J1815-195 shows a single r-band point potentially in outburst. The position of this brightening corresponds to source 3 in atel \#15479 \citep{deMartino:2022:optMAXIJ1826}}
    \label{fig:xrayonly}
\end{figure}

\end{document}